\documentclass{SolarPhysics}    
\usepackage[optionalrh]{spr-sola-addons}
\usepackage{epsfig}
\usepackage{url}
\usepackage{fixltx2e}
\usepackage[usenames]{color}

\newcommand{\ang}{\AA\ }
\newcommand{\gapprox}{\lower.4ex\hbox{$\;\buildrel >\over{\scriptstyle\sim}\;$}}
\newcommand{\lapprox}{\lower.4ex\hbox{$\;\buildrel <\over{\scriptstyle\sim}\;$}}
\newcommand{\arcsec}{\hbox{$^{\prime\prime}$}}

\def\ang{\AA}

\begin{document}
\begin{article}
\begin{opening}
\title{ 	Benchmark Test 
		of Differential Emission Measure Codes 
		and Multi-Thermal Energies in Solar Active Regions }

\author{Markus J. Aschwanden$^1$ \sep 
		Paul Boerner$^1$ \sep 
		Amir Caspi$^2$ \sep 
		James M. McTiernan$^3$ \sep
		Daniel Ryan$^4$ \sep
		Harry P. Warren$^5$} 

\runningauthor{M.J. Aschwanden \textit{et al.}}
\runningtitle{DEM Benchmark Test}

\institute{$^1)$Solar and Astrophysics Laboratory,
		Lockheed Martin Advanced Technology Center, 
        	Dept. A021S, Bldg.252, 3251 Hanover St., 
		Palo Alto, CA 94304, USA; 
        	(e-mail: \url{aschwanden@lmsal.com})}

\institute{$^2)$Planetary Science Directorate,
		Southwest Research Institute;
                Boulder, CO 80302, USA;
                (e-mail: \url{amir@boulder.swri.edu})}

\institute{$^3)$Space Sciences Laboratory,
		University of California,
		Berkeley, CA 94720, USA;
		(e-mail: \url{jimm@ssl.berkeley.edu})}

\institute{$^4)$Royal Observatory of Belgium,
		Solar-Terrestrial Centre for Excellence,
		Avenue Circulaire 3,
		1180 Uccle, Brussels, Belgium;
                (e-mail: \url{ryand5@tcd.ie})}

\institute{$^5)$Space Science Division,
                Naval Research Laboratory,
                Washington, DC 20375, USA;
                (e-mail: \url{harry.warren@nrl.navy.mil})}

\date{Received 15 Dec 2014; Revised ...; Accepted ...}

\begin{abstract}
We compare the ability of 11 Differential Emission Measure (DEM) 
forward-fitting and inversion methods to constrain the properties of
active regions and solar flares by simulating synthetic data using 
the instrumental response functions of SDO/AIA, SDO/EVE, RHESSI, 
and GOES/XRS. The codes include the 
single-Gaussian DEM, a bi-Gaussian DEM, a fixed-Gaussian DEM, 
a linear spline DEM, the spatial synthesis DEM, the Monte-Carlo 
Markov chain DEM, the regularized DEM inversion, the Hinode/XRT 
method, a polynomial spline DEM, an EVE+GOES, and 
an EVE+RHESSI method. Averaging the results from all 11 DEM methods,
we find the following accuracies in the inversion of physical parameters:
the EM-weighted temperature $T_w^{fit}/T_w^{sim}=0.9\pm0.1$, 
the peak emission measure $EM_p^{fit}/EM_p^{sim}=0.6\pm0.2$,
the total emission measure $EM_t^{fit}/EM_t^{sim}=0.8\pm0.3$, and
the multi-thermal energies $E_{th}^{fit}/EM_{th}^{sim}=1.2\pm0.4$.
We find that the AIA spatial synthesis, the EVE+GOES, and the EVE+RHESSI
method yield the most accurate results.
\end{abstract}


\end{opening}

\section{	INTRODUCTION 					}

A benchmark test of {\sl differential emission measure (DEM)} analysis
methods is timely since the current capabilities from the {\sl 
Atmospheric Imaging Assembly} (AIA; Lemen et al.~2012) and the
{\sl EUV Variability Experiment} (EVE; Woods et al. 2012) onboard
the {\sl Solar Dynamics Observatory} (SDO; Pesnell et al.~2012),
combined with simultaneous data from the {\sl Reuven Ramaty High 
Energy Solar Spectroscopic Imager} (RHESSI; Lin et al.~2002) and 
the X-ray Sensor (XRS; Garcia 1994) onboard the {\sl Geostationary 
Operational Environmental Satellite} (GOES) series,
now offer unprecedented opportunities to design
DEM algorithms that take advantage of comprehensive temperature 
coverage, high spectral resolution, and high spatial resolution 
in resolving and discriminating complex temperature structures 
in the solar corona. Another motivation for this study is the 
calculation of accurate multi-thermal energies, compared with 
the previously used isothermal approximations, which is the 
subject of a recent major study on the global energetics of solar flare 
and coronal mass ejection (CME) events (Aschwanden et al.~2015). 

The concept of DEM distributions and the ill-posed problem of the 
inversion from the observed radiation of optically thin thermal 
emission (produced by bremsstrahlung [free-free], radiative 
recombination [free-bound] emission, and even [bound-bound] emission, 
has been recognized early on
(Craig and Brown 1976; Judge, Hubeny, and Brown 1997). It was pointed
out that systematic errors resulting from incomplete calculations
of atomic excitation levels and from data noise represent a fundamental
limitation in DEM inversion (Judge 2010). The inversion of simple
synthetic Gaussian or rectangular DEMs was tested with the
{\sl emission measure loci} method, which can retrieve the
temperature width of a single-peaked DEM (Landi and Klimchuk 2010).  
Tests of isothermal DEMs with the {\sl Monte Carlo Markov chain
(MC)} method (Kashyap and Drake 1998), including data noise 
and uncertainties in the
atomic data, revealed that the MC method cannot resolve
isothermal plasmas better than $\Delta log(T) \approx 0.05$,
and that two isothermal components can not be resolved better 
than $\Delta log(T) \approx 0.2$ (Landi, Reale, and Testa 2012).
Tests on synthetic single-Gaussian and multi-Gaussian DEMs with
the {\sl regularized inversion} technique yielded uncertainties
of $\Delta log(T) \approx 0.1-0.5$ and a valid range of the
retrieved DEM down to about a level of $\gapprox 1\%$ of the DEM peak 
emission measure (Hannah and Kontar 2012). 

Most of the previously employed DEM methods use spatially-integrated
EUV and/or soft X-ray fluxes as constraints, which, for SDO/AIA data, 
yields typically 6 flux values that are used in the inversion of
the DEM function (e.g., Aschwanden and Shimizu 2013). For SDO/EVE, 
a spatially-integrated spectrometer with $\approx 1$ \ang\ spectral 
resolution, the data include many tens of spectral lines across 
hundreds of spectral bins (Warren et al.~2013). However, these 
EUV-based methods are typically poorly constrained at high 
temperatures (above $\log(T)~\approx 7.3$) due to the relative 
lack of EUV lines from solar-abundant ions sensitive to these 
temperatures (e.g., Winebarger et al.~2012).
Some more recent DEM methods extend the high-temperature 
coverage by including GOES/XRS (Warren et al.~2013; Warren 2014) 
and/or RHESSI (Caspi et al.~2014b; Inglis and Christe 2014) X-ray data, 
enabling complete coverage of coronal temperatures from 
$\log(T)~\gapprox 6.3$. Fluxes that are spatially integrated over 
a substantial area of a flare or active region naturally contain 
many temperatures and thus can have arbitrarily complex broadband DEMs. 

A novel method consists of
subdividing the observed space into small areas (called
macro-pixels or cells), down to the pixel size of the image, which 
are more likely to encompass a narrower and simpler temperature 
distribution due to the smaller number of bright structures that
are intersected, and then to perform a DEM reconstruction in every
macro-pixel, while the total DEM of the entire flare area or active
region can then simply be added together. Such a {\sl spatial-synthesis
method (SS)} has been developed for SDO/AIA recently (Aschwanden
et al.~2013), and further developments are underway (Cheung et al.~2015).

In this paper we conduct a benchmark test of a set of 11 different DEM
forward-fitting and inversion schemes, which include both types of
total-flux and spatially-synthesized DEM methods, in order to compare and 
contrast their accuracy and precision. We perform this test by 
creating realistic, but synthetic, 2D projected images of flaring loops 
at all wavelengths using a numerical simulation code; the simulated 
images are then used to construct spatially-integrated DEMs using the 
11 schemes, and are then compared to the true DEM derived from the 
model input.

\section{	DATA SIMULATION  				}

\subsection{	Simulated Differential Emission Measure Maps 	}

We simulate synthetic data of differential emission measure maps $EM(x,y,T)$
as a function of the temperature $T$, in the form of spatial images in the 
$(x,y)$ plane, sampled in various temperature intervals $[T, T+\Delta T]$ 
that cover the EUV and soft X-ray wavelength range in a temperature 
range of $T_e=10^5-10^8$ K. The simulated data aim to mimic loop 
arcades of large flares with a complex multi-temperature structure.

\begin{figure}
\centerline{\includegraphics[width=1.0\textwidth]{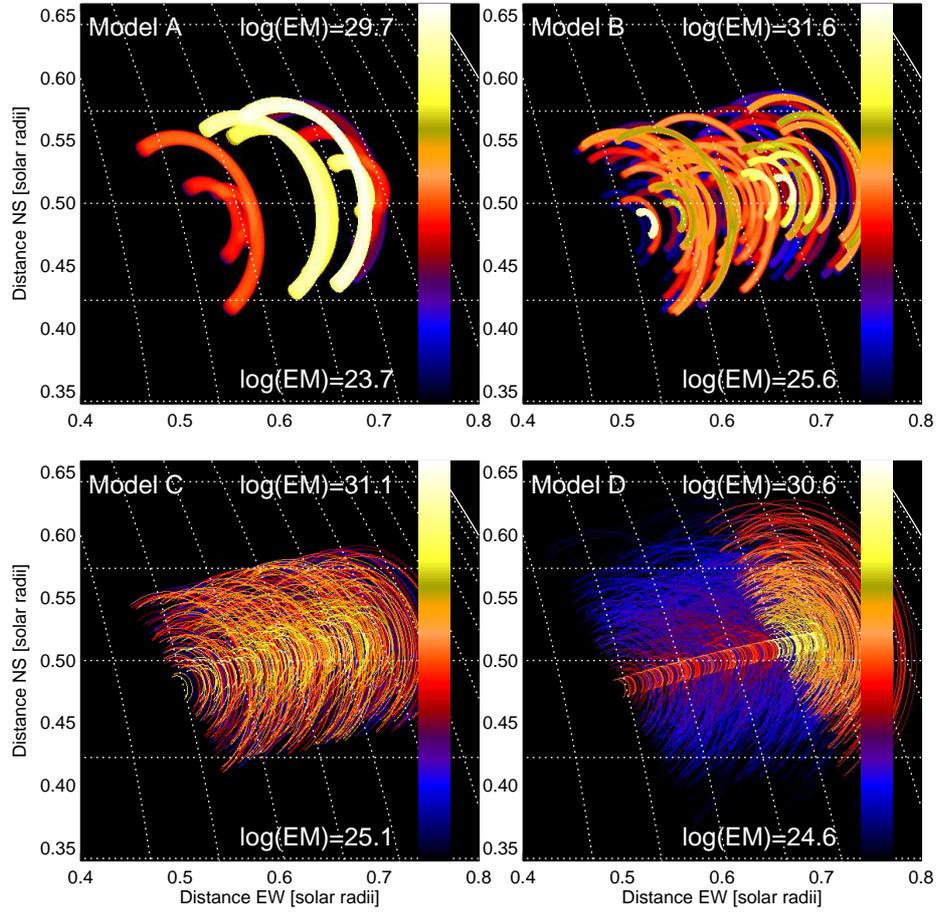}}
\caption{Simulated total emission measure maps $EM(x,y)$ of the four models:
(A) the monolithic scenario with 10 loops with a width of 11 pixels;
(B) the intermediate scenario with 100 loops with a width of 5 pixels;
(C) the nanoflare scenario with 1000 loops with a width of 1 pixel, and
(D) a dual-temperature population model with a width of 1 pixel.
The lower and upper limits of $log(EM)$ of the color bar is indicated
in each panel. The x-axis and y-axis represents the distance from 
Sun center in units of solar radii.}
\end{figure}

The synthetic data cover a $512 \times 512$ pixel image with a pixel
size of $\Delta x = 0.6\arcsec \approx 435$ km, corresponding
to the SDO/AIA pixel scale. The simulated images
contain $n_{loop}=$ 10, 100, or 1000 
semi-circular flare loops that are randomly distributed along a 
neutral line, which is centered at a given heliographic location 
(at longitude $l_0=45^\circ$ and latitude $b_0=30^\circ$) 
and has a length of $L_0=200 \Delta x \approx 0.125$ solar radii, and 
is oriented with an azimuthal angle $\alpha=10^\circ$
with respect to the East-West direction (x-axis); see an example
in Fig.~1. 
The loops have a half-length that is randomly distributed 
within a range of $L=(20-200) \Delta x \approx 10-100$ Mm, 
and have an apex temperature that is randomly distributed within 
a logarithmic range of $log(T_{max})=6.0 - 7.3$ K.
The temperature profile
$T(s)$ along the loops follows the hydrostatic approximation
(Aschwanden and Tsiklauri 2009),
\begin{equation}
	T(s) = T_{max} \left[ \left( {s \over L} \right)
	\left( 2 - {s \over L} \right) \right]^{2/7} \ , 
\end{equation}
with a minimum temperature of $T_{min}=10^5$ K at the footpoints.
The electron density is uniform along the loop and follows
the Rosner-Tucker-Vaiana (RTV) scaling law (Rosner et al.~1978), 
\begin{equation}
	n_e= (8.4 \times 10^5) T_{max}^2 L^{-1} \ ,
\end{equation}
which covers a range of $n_e \approx 1.0 \times 10^8 - 
1.5 \times 10^{11}$ cm$^{-3}$ here. 

We simulate 4 sets of images: Model (A) with a small number of $n=10$
thick loops (with a radius of $r=11$ pixel) that may be typical 
for spatially resolved monolithic loops (e.g., Aschwanden et al.~2007);
Model (B) with a medium number of $n=100$ loops and an intermediate radius
of $r=5$ pixels; Model (C) with a large number of $n=1000$ very
slender loop strands with a radius of $r=1$ pixel that is typical
for unresolved nanoflare loop strands (e.g., Scullion et al.~2014); 
and Model (D) that contains 
$n=1000$ loops in two different temperature regimes that form a 
double-peaked DEM and falls off less steeply at the high-temperature
tail of the DEM, while models A, B, and C have a sharp cutoff. 

The four models are designed to represent realistic simulations
of active regions in the solar corona, regarding spatial structures,
temperature distributions, and realistic electron densities as
obtained from the RTV scaling law (Rosner et al.~1978). 
In addition, these four models represent different thermo-spatial patterns,
being essentially isothermal in a single pixel with fully resolved
loops (Model A), or inherently multi-thermal with spatially unresolved
strands (Model C), which cover both extremes (of fully resolved or 
spatially unresolved temperature structures) to test the capabilities
of the different DEM inversion codes. The thermo-spatial pattern
matters mostly for pixel-based DEM inversion codes (such as the spatial
synthesis code; Section 4.5), while all other codes tested here do not
use any spatial information in the DEM inversion. Therefore, we will
learn whether or not the inclusion of spatial information affects the
accuracy of DEM reconstructions. One new DEM inversion code that uses 
spatial information also has been designed recently (Cheung et al.~2015). 

Each individual loop is 
uniformly filled with space-filling voxels $(x,y,z)$ that have a volume 
$(\Delta x)^3$, a unique electron temperature $T(x,y,z)$, and electron density 
$n_e(x,y,z)$. The pixelized data cube $EM(x,y,z,T)$ is then
integrated along the line-of-sight in direction of the z-axis
in order to produce differential emission measure maps in each 
temperature interval,
\begin{equation}
	EM(x,y,T)= \int n_e^2(x,y,z,T) dz \ ,
\end{equation}
which has the units of [cm$^{-5}$ K$^{-1}$]. Simulated total emission measure
maps, integrated over the temperature range $T$ with units of [cm$^{-5}$],
\begin{equation}
	EM(x,y) = \int EM(x,y,T) dT \ ,
\end{equation}
are shown for all 4 cases in Fig.~1. 
The temperature is discretized in a logarithmic range of 
$log(T_e)=5.00, 5.05, ..., 8.00$ in 61 equidistant logarithmic
steps with a width of $\Delta log(T)=0.05$ each. 
A total differential emission measure distribution $EM_{tot}$ 
integrated over the entire volume is defined by
\begin{equation}
	EM_{tot}(T) = \int\int EM(x,y,T) \ dx dy \ ,
\end{equation}
and has the units of [cm$^{-3}$ K$^{-1}$]. 

\subsection{	Simulation of AIA Flux Maps 			}

The differential emission measure maps $EM(x,y,T)$ are defined in an
instrument -independent way. In order to simulate observables, 
the simulated theoretical differential emission measure $EM(x,y,T)$ is then 
convolved with the instrumental response function $R_{\lambda}(T)$
(in units of [DN s$^{-1}$ cm$^5$] per pixel)
of a particular temperature filter, in order to obtain the
expected flux maps $f_{\lambda}(x,y)$ (in units of DN/s), 
\begin{equation}
        f_{\lambda}(x,y) = \int EM(x,y,T) \ R_{\lambda}(T) \ dT
        = \sum_{k=1}^{n_T} EM(x,y,T_k)\ R_{\lambda}(T_k) 
	\Delta T_k \ ,
\end{equation}
where $n_T$ is the number of temperature filters, and
$f_{\lambda}$ are the observed fluxes 
in the coronally-domainted EUV wavelengths 
$\lambda$ = 94, 131, 171, 193, 211, 335 \ang,
in the case of the SDO/AIA instrument.
For the temperature integration we are using the discretized
temperature intervals $\Delta log(T_k)$, which are chosen
in equidistant bins with a width of 
$\Delta log(T_k) = log(T_{k+1})-log(T_k) = 0.05$.
This way we obtain 6 simulated AIA images for which the
differential emission measure distribution $EM(x,y,T)$ is exactly known
from the theoretical model. We can then use these flux maps 
$f_{\lambda}(x,y)$, $\lambda=1,...,6$ to test the various DEM 
inversion codes and methods.  An example of 6 simulated AIA flux maps 
is shown in Fig.~2 (Model B).

\begin{figure}
\centerline{\includegraphics[width=1.0\textwidth]{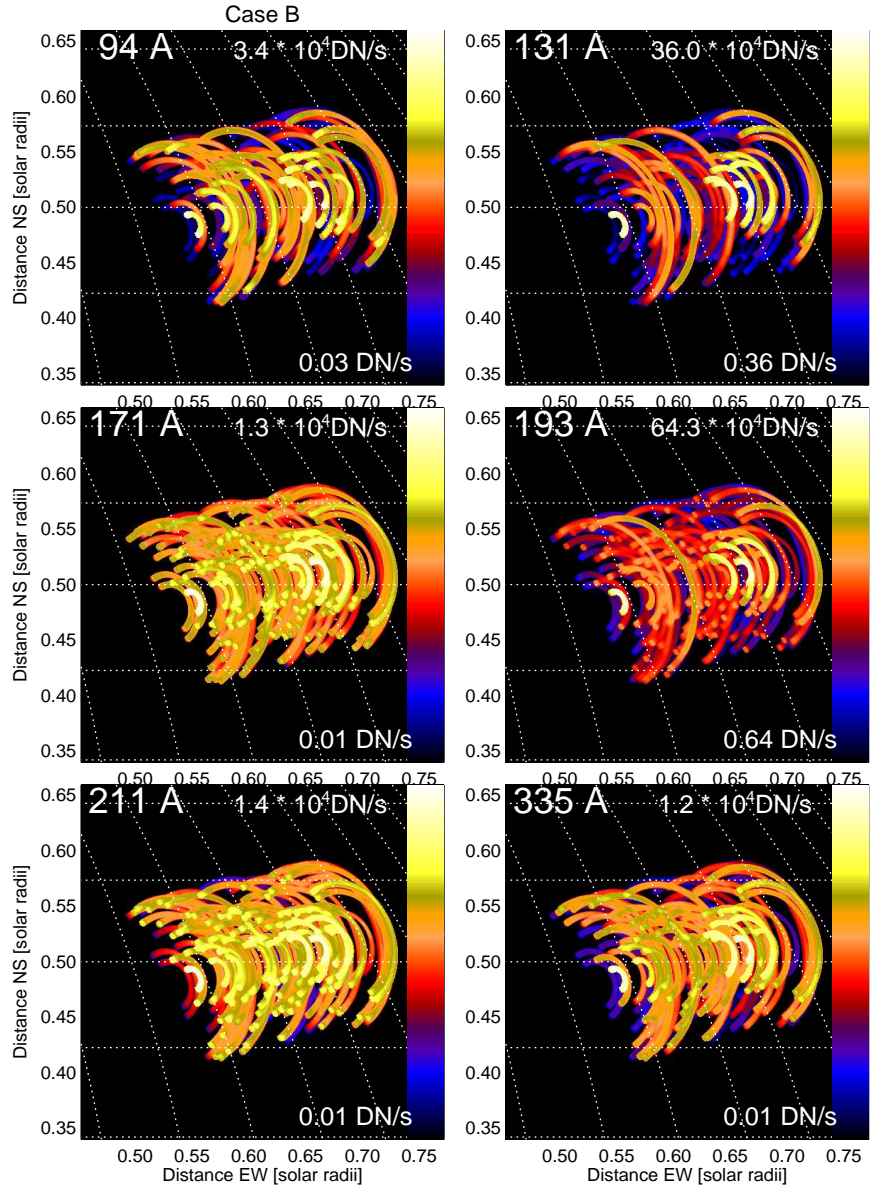}}
\caption{Simulated AIA flux maps $F_{\lambda}(x,y)$ for the 6
AIA wavelengths. The simulation of model B includes 100 flare
loops with a width of 5 pixels in a temperature range of
$T_e=1-20$ MK. The lower and upper flux limits of the color bar
are indicated in each panel. The x-axis and y-axis
represents the distance from Sun center in units of solar radii.}
\end{figure}

\subsection{	Multi-Thermal Energy				}

The multi-thermal energy, which is the integral of all thermal energies
integrated over each temperature range $\Delta T_k$, is defined as
\begin{equation}
        E_{th} = \sum_k 3 k_B V^{1/2}\ T_k\ EM_k^{1/2} \ .
\end{equation}
This expression can substantially deviate from the isothermal approximation
(independent of the DEM inversion method),
\begin{equation}
         E_{th,iso} = 3 n_p k_B T_p V
                     = 3 k_B T_p \sqrt{EM_p \ V} \ ,
\end{equation}
which assumes a narrow delta-like DEM distribution that can be characterized
by the DEM peak temperature $T_p$ and DEM peak emission measure $EM_p$.
Separate calculations of isothermal and multi-thermal energies based on
the same DEM fitting method in $\approx 400$ M- and X-class flares has 
shown that the multi-thermal energy exceeds the isothermal energy by a 
factor of $\sim 14$ in the statistical average (Aschwanden et al.~2015).
A comparison with these multi-thermal energies shows that our 4 models
have conditions that are typical for active regions, rather than for 
large flares.

\section{	INSTRUMENTAL RESPONSE FUNCTIONS  	}

\subsection{	The SDO/AIA Response Functions 		}

The AIA instrument onboard SDO
started observations on 29 March 2010 and has produced since then essentially
continuous data of the full Sun with four $4096 \times 4096$ detectors
with a pixel scale of $0.6\arcsec$, corresponding to an effective
spatial resolution of $\approx 1.5\arcsec$ (Lemen et al.~2012). 
AIA contains ten different
wavelength channels, three in white light and UV, and seven EUV channels,
of which six wavelengths (131, 171, 193, 211, 335, 94 \ang ) are centered on 
strong iron lines (Fe {\sc viii}, {\sc ix}, {\sc xii}, {\sc xiv}, {\sc xvi}, 
{\sc xviii}), covering the coronal range from $T\approx 0.6$ MK to 
$\gapprox 16$ MK.  AIA records a full set of near-simultaneous images 
in each temperature filter with a fixed cadence of 12 s. 
Instrumental descriptions can be found in Lemen et al.~(2012) 
and Boerner et al.~(2012, 2014). The nominal AIA response functions 
$R_{\lambda}(T)$ are shown in Fig.~3. We use the currently
available calibration, which was updated with improved atomic emissivities
according to the CHIANTI Version 7 code, distributed in the 
{\sl Solar SoftWare (SSW)} on 2012 February 13. Although
the response functions of the AIA channels (with a full width of 
$\log{(\Delta T)} \approx 0.2$) are relatively narrowband (compared
with GOES or Yohkoh/SXT), we have to be aware that this intrinsic
temperature width represents an ultimate lower limit for resolving
multiple temperature structures, and thus constitutes a bias in the
inversion of narrrower (more isothermal) temperature structures
in the DEM inversion, as previously verified (e.g., Kashyap and Drake 
1998; Landi et al., 2012). An additional caveat of widely distributed
DEMs is that the intensity is completely smoothed out, regardless
of the shape of the response function. Although AIA has the ability to
reconstruct simple DEM forms, the greater the width of the plasma DEM is,
the lower is the accuracy and precision in the determination of the DEM
parameters, which is a fundamental limitation (Guennou et al.~2012).

Another uncertainty in DEM reconstructions comes from atomic physics,
such as the atomic line emissivities, the assumption of ionization
equilibrium, and variations in elemental abundances between the
chromosphere and corona. The AIA response functions were calculated
based on the latest version (7.1) of the CHIANTI code. While the line
intensities in the wavelength range of 170--350 \ang\ included in
CHIANTI are considered to be satisfactory, the CHIANTI Version 7.1 code
appears to underpredict the observed intensities in the 50--150 \ang\
wavelength range by factors of 2-6 (Boerner et al.~2014), which affects mostly 
the 94 and 131 \ang\ channels (Aschwanden and Boerner 2011; Teriaca, 
Warren, and Curdt, 2012), a fact that was also corroborated with 
measurements of spectra from the star Procyon taken by Chandra's LETG 
(Testa, Drake, and Landi 2012b). Regarding elemental abundance variations, 
AIA has been designed to consist of 6 coronal wavelength channels that 
are all sensitive to iron lines, and thus the elemental abundance of 
iron, which exhibits the largest first-ionization potential (FIP) 
effect, largely cancels out in the combination of the six coronal AIA 
channels. However, although the shape of the reconstructed DEM 
is not much affected when only iron lines are used, the magnitude of
the DEM still depends on the absolutie value of the iron abundance,
which may have a substantially larger error than 10\%. In our 
benchmark study we choose to neglect the atomic physics uncertainties, 
which makes the expected cancellation of the FIP effect less certain.

The calibration uncertainties of the AIA response function have 
been estimated to be 25\% in absolute terms, using a comparison of
full-disk-integrated fluxes with SDO/EVE, but reduce to $\lapprox 10\%$
after application of the cross-calibration (Boerner et al.~2014).
The residual ratios left from fitting the AIA /EVE flux ratios are
shown in Fig.~2 of Boerner et al.~(2014) and can also be obtained
from the IDL {\sl Solar SoftWare (SSW)} procedure 
{\sl aia$\_$bp$\_$read$\_$error$\_$table}. The individual uncertainties in 
the different AIA channels amount to:
8.7\% ( 94 \ang ), 5.1\% (131 \ang ), 1.9\% (171 \ang ),
1.4\% (193 \ang ), 1.9\% (211 \ang ), 2.3\% (304 \ang ),
9.7\% (335 \ang ), so we conservatively adopt 10\% as an upper limit.

\begin{figure}
\centerline{\includegraphics[width=1.0\textwidth]{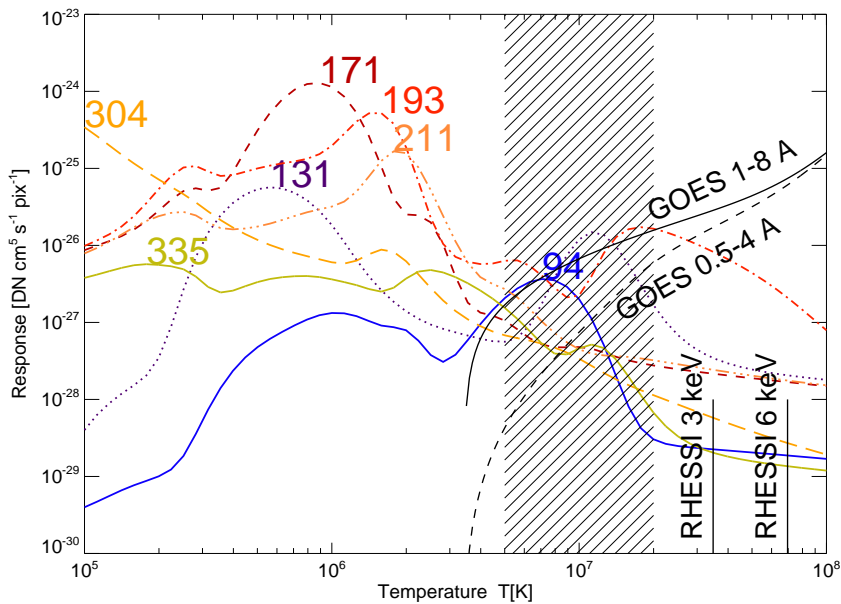}}
\caption{Temperature-response functions for the seven coronal EUV
channels of the {\sl Atmospheric Imaging Assembly (AIA)} onboard
the {\sl Solar Dynamics Observatory (SDO)}, according to the status 
of Dec 2012.  The GOES/XRS 1-8 \ang \ and 0.5-4 \ang \ response 
is also shown (in arbitrary flux units),
as well as the thermal energy of the lowest fittable RHESSI
channels at 3 keV and 6 keV. The approximate peak temperature range of large
flares ($T_p \approx 5-20$ MK) is indicated with a hatched area.}
\end{figure}

\begin{figure}
\centerline{\includegraphics[width=1.0\textwidth]{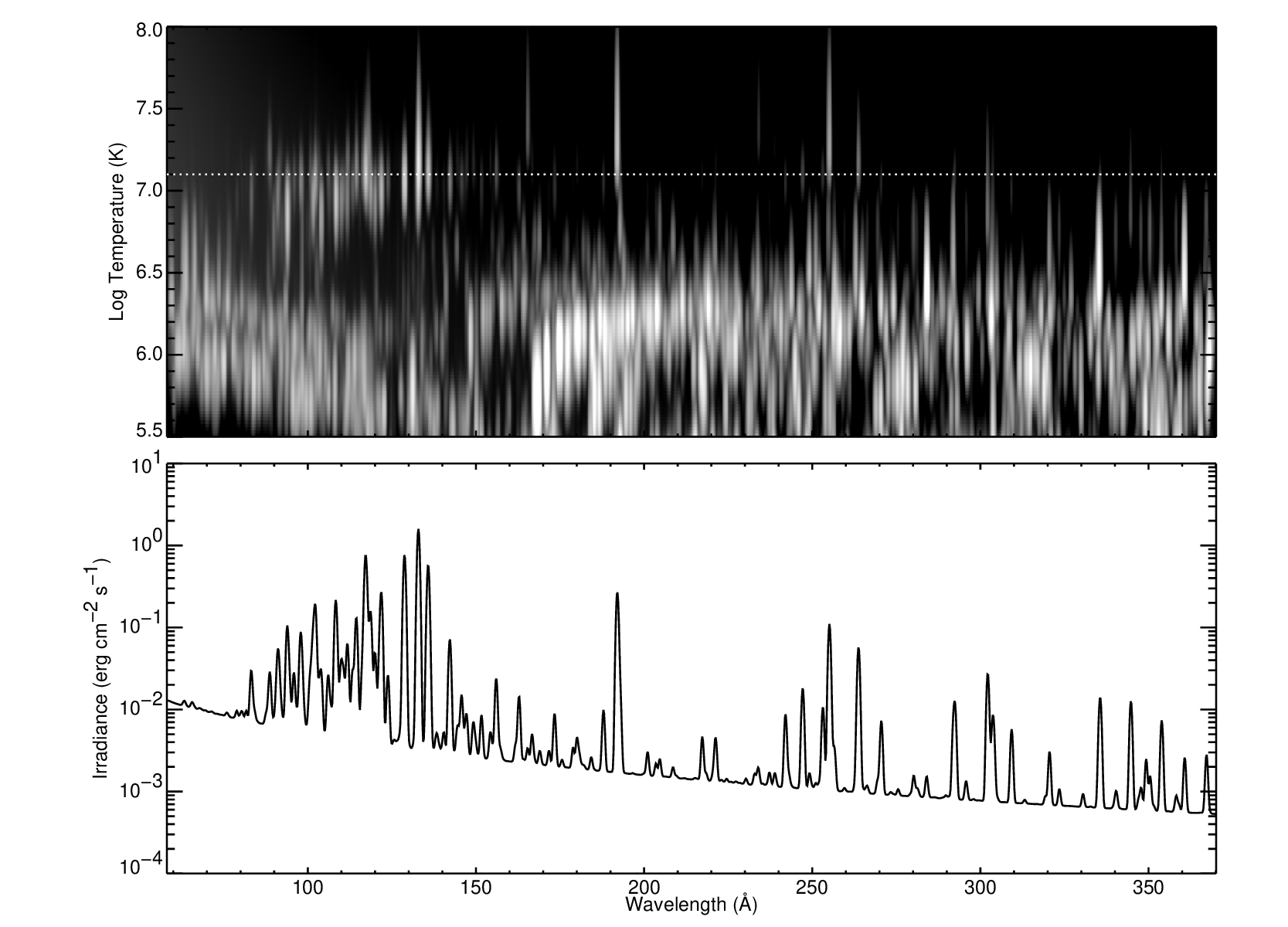}}
\caption{The SDO/EVE response function $R_{EVE}(\lambda, T)$, 
represented here as the CHIANTI emissivity convolved with the 
EVE spectral resolution, as a function of the
wavelength $\lambda$ (horizontal axis) and temperature (vertical 
axis in top panel), in the range of $\lambda=60-350$ \ang\ and 
$log(T_e)=5.5-8.0$.  (bottom panel) An isothermal spectrum
at a temperature of $log(T_e)=7.1$ and an emission measure of 
$10^{51}$\,cm$^{3}$ is shown.  The temperature is indicated by 
a dotted line in the upper panel).}
\end{figure}

\subsection{	The SDO/EVE Response Function 		}

The EVE onboard SDO measures the solar irradiance at many EUV wavelengths. 
Here we use observations from the {\sl Multiple EUV Grating
Spectrograph A (MEGS-A)}, a grazing incidence spectrograph that 
operates in the 50-370 \ang\ wavelength range, has a spectral 
resolution of $\sim 1$ \ang , and an observing cadence of 10 s.  
The observed emission at these wavelengths is sensitive to 
temperatures ranging from the upper chromosphere (He {\sc II} 
304 \ang ) to about 25\,MK (Fe {\sc XXIV} 192 and 255 \ang). 
Since the peak temperature sensitivity of the MEGS-A is below 
the peak temperature observed in many flares, it is useful to 
combine EVE observations with observations from other
instruments that are sensitive to higher temperature emission 
(e.g., GOES/XRS, RHESSI).

The modest spectral resolution of the MEGS-A means that while 
many spectral features can be easily identified, many are blended 
and reliably calculating the fluxes of individual emission lines is
difficult. Our strategy is to compute theoretical spectra from CHIANTI, 
convolve them with the EVE spectral resolution, 
and to compare these spectra directly 
with the observations in selected wavelength ranges. The calculated 
spectra used for this work are shown in Figure 4. Further details 
are given in Warren et al.~(2013) and Caspi et al.~(2014b).

We note that since the spectral information is preserved in 
the treatment of the EVE observations, the application of random 
perturbations affects these data differently than other broadband 
measurements. For the broadband measurements the integrated 
intensity is perturbed. With the EVE data we are perturbing the 
intensity in each spectral pixel and thus the fluctuations tend to
 average out, leaving the intensity in each spectral feature 
largely unchanged. A more realistic treatment would be to smoothly 
vary the effective areas that are used to convert the observed 
count rates to absolute intensities. Our expectation is that the 
uncertainty in the calibration is dominated by systematic trends 
and not pixel-to-pixel noise. Such a procedure would be relatively 
easy to implement, but is beyond the scope of this work.

One subtlety of analyzing the actual EVE observations is that 
unidentified emission lines make significant contributions to some 
wavelength ranges and the calculated spectra cannot be matched
to the observations in these regions (Warren 2005; Testa et al.~2012b). 
This can be addressed by subtracting a pre-flare spectrum from the 
observed spectra during the flare (e.g., Warren et al.~2013;
Caspi et al.~2014b). For this work, which considers 
theoretical spectra, we do not apply background subtraction since 
no background is simulated in the synthesized input.

\subsection{	The RHESSI Response Function 		}

RHESSI observes solar photons from $\sim$ 3 keV to $\sim$ 17 MeV 
(Lin et al.~2002), using a set of nine cryogenically cooled coaxial 
germanium detectors to achieve $\lapprox$1~keV FWHM spectral resolution 
(Smith et al. 2002).  RHESSI is also capable of imaging X-ray and 
gamma-ray sources through a Fourier method (Hurford et al. 2002), 
but we do not use the imaging capabilities for this work.

Since flare temperatures of $T_e \approx 5-50$ MK correspond to thermal
electron energies of $E_{th} = k_B T_e \approx 0.4-4.4$ keV (Fig.~3), RHESSI 
is sensitive to the high-temperature tails of flare DEMs, whose emission 
typically dominates the RHESSI spectrum in the $3-20$ keV energy range.
While most of the thermal emission detected in this range is produced
by the free-free (bremsstrahlung) and free-bound (radiative recombination) 
continuum processes, RHESSI also observes the line emission of two 
iron-dominated line complexes at $\sim 6.7$ (Fe) and $\sim 8$ (Fe/Ni) keV 
(Phillips 2004; Phillips et al.~2006; Phillips 2008; Caspi and Lin 2010).

When multiple thermal components are present, fitting of RHESSI thermal 
spectra in isolation has a bias towards higher temperatures because of 
RHESSI's exponentially-increasing temperature sensitivity (McTiernan 2009). 
For intense flares where the DEM tail extends to very high temperatures, 
an isothermal or delta-function DEM analysis often yields a ``super-hot''
component in the temperature range of $T_e \approx 20-50$ MK
(e.g., Caspi and Lin 2010; Caspi et al.~2014a), which is significantly 
higher than the DEM peak temperature (with a typical range of
$T_e \approx 10-20$ MK), a bias that was also quantitatively
investigated in Ryan et al.~(2014). Improved temperature diagnostics
can be obtained by combining RHESSI high-temperature spectra with the 
broadband EVE irradiance spectra that are sensitive to the 
lower-temperature emission, a method that has been successfully applied 
to two X-class flare events so far (Caspi et al.~2014b).

\subsection{	The GOES Response Function 		}

The GOES X-Ray Sensor (XRS) measures spatially integrated broadband solar 
X-ray radiation in two overlapping passbands: 1-8 \ang\ and 0.5-4 \ang . 
In typical solar conditions, it is most sensitive to coronal emission 
from temperatures of $\sim 4-40$ MK. This is ideal for examining the bulk 
thermal emission of M- and X-class flares which have typical peak 
temperatures of $10-25$ MK (Ryan et al.~2012). The XRS detectors 
comprise two ion chambers, one for each channel. The output currents 
of the ion chambers are related to the incident X-ray flux via 
wavelength-dependent transfer functions (Garcia 1994; Hanser and 
Sellers 1996). The GOES/XRS response function is calculated based on
the CHIANTI code assuming photospheric (or coronal) elemental abundances 
and ionization equilibrium (e.g. Feldman et al.~1992; Mazzotta et al.~1998).  
Normally when analyzing GOES/XRS observations, the DEM is assumed to be 
isothermal and is then derived from the ratio of the short and long channel 
fluxes (Thomas et al.~1985; Garcia 1994; White et al.~2005). This 
limitation is given because the two-channel data can only constrain a 
single temperature with a filter-ratio technique. The isothermal 
approximation, however, has a temperature bias for multi-thermal DEM 
distributions (see Eq.~(7) in Ryan et al.~2014). The two-channel data 
alone also cannot strongly constrain broad DEM solutions of 
forward-fitting techniques, resulting in a wide range of non-unique 
solutions. Combining GOES/XRS with other instruments, however, such as 
with SDO/EVE (Warren et al. 2013) as used in this work, provides the 
necessary constraints to yield more realistic multi-thermal DEMs. 

\begin{figure}
\centerline{\includegraphics[width=1.0\textwidth]{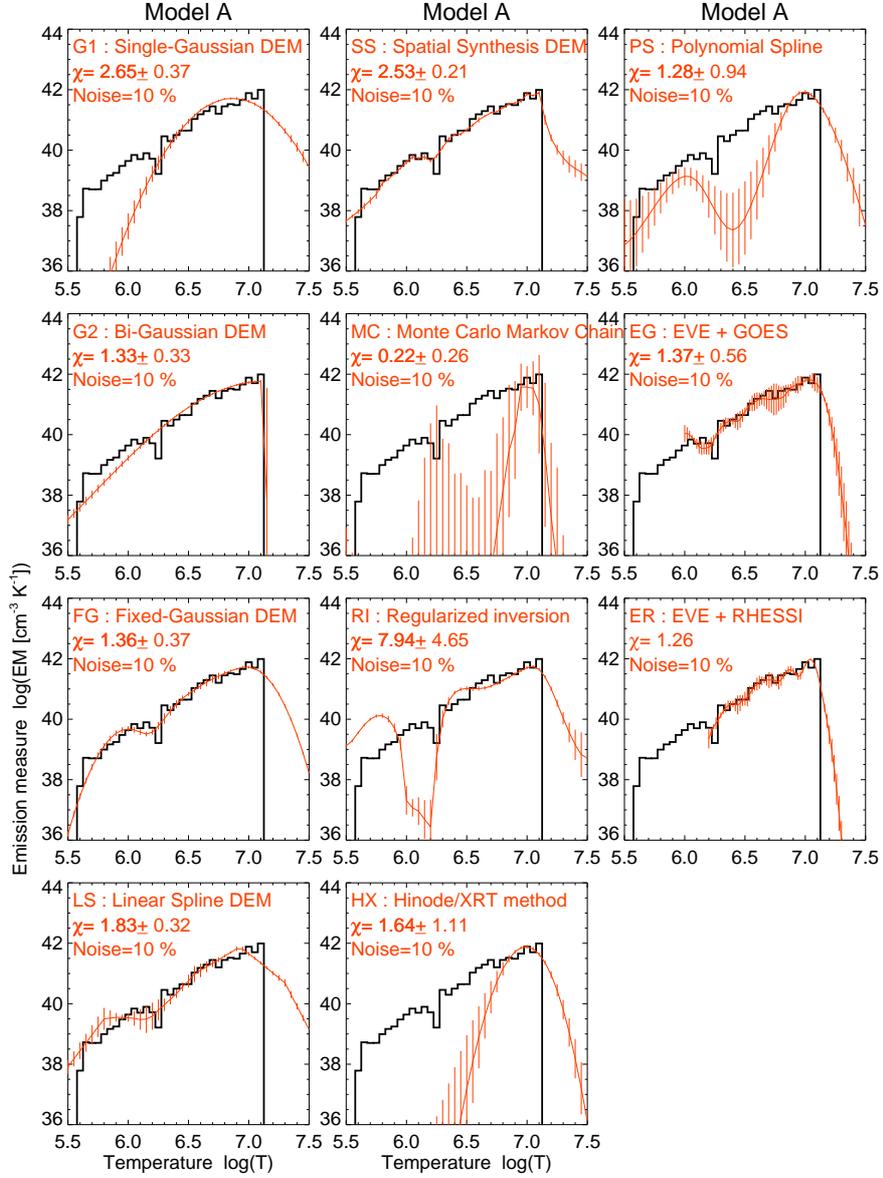}}
\caption{DEM inversions of monolithic loop model A, using
11 different DEM forward-fitting or inversion methods.
The simulated DEM is indicated with a black histogram,
while the average of 30 DEM inversions (by adding 10\% 
noise to the simulated fluxes) are indicated with red
curves, including error bars. The reduced $\chi^2$ is
computed from the average of the 30 runs. The x-axis 
represents the logarithm of the temperature in units of K.}
\end{figure}

\begin{figure}
\centerline{\includegraphics[width=1.0\textwidth]{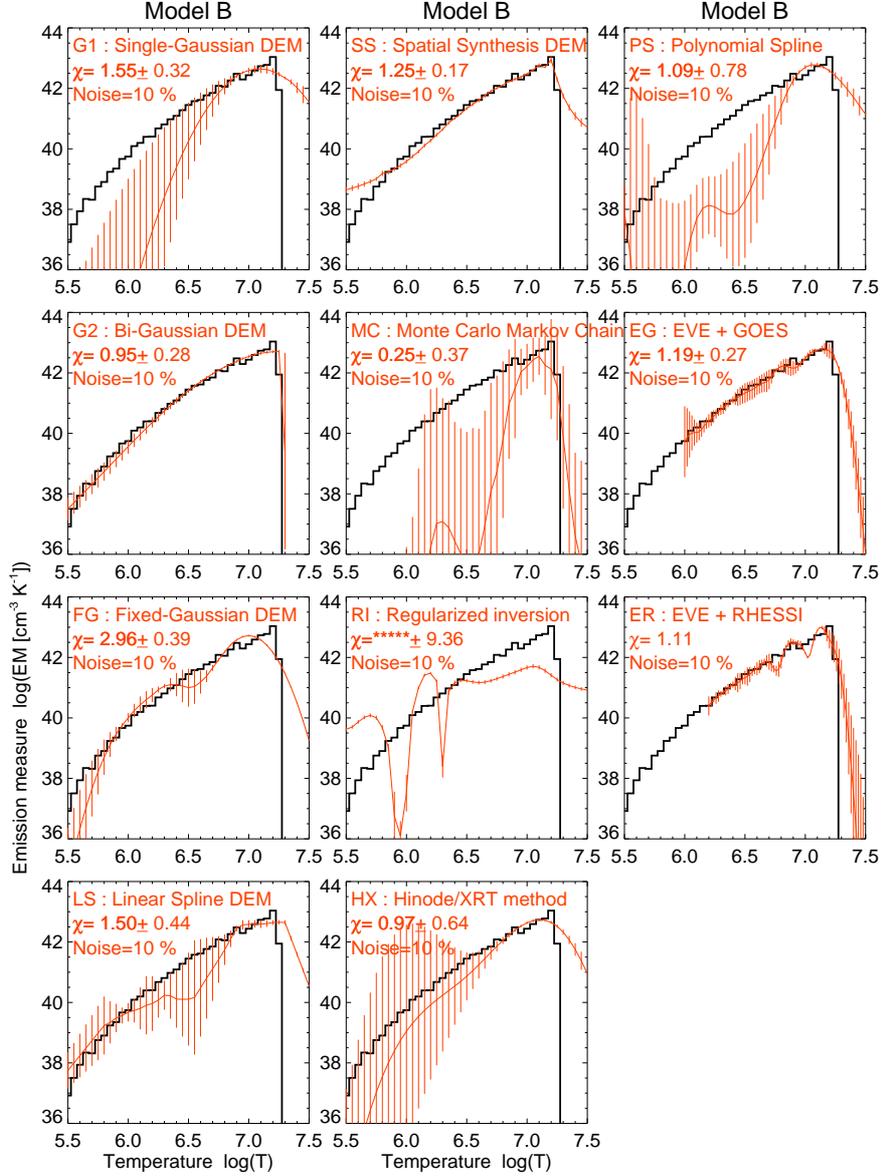}}
\caption{DEM inversions for thin loop model B, using
11 different DEM methods. Otherwise representation similar
to Fig.~5.}
\end{figure}

\begin{figure}
\centerline{\includegraphics[width=1.0\textwidth]{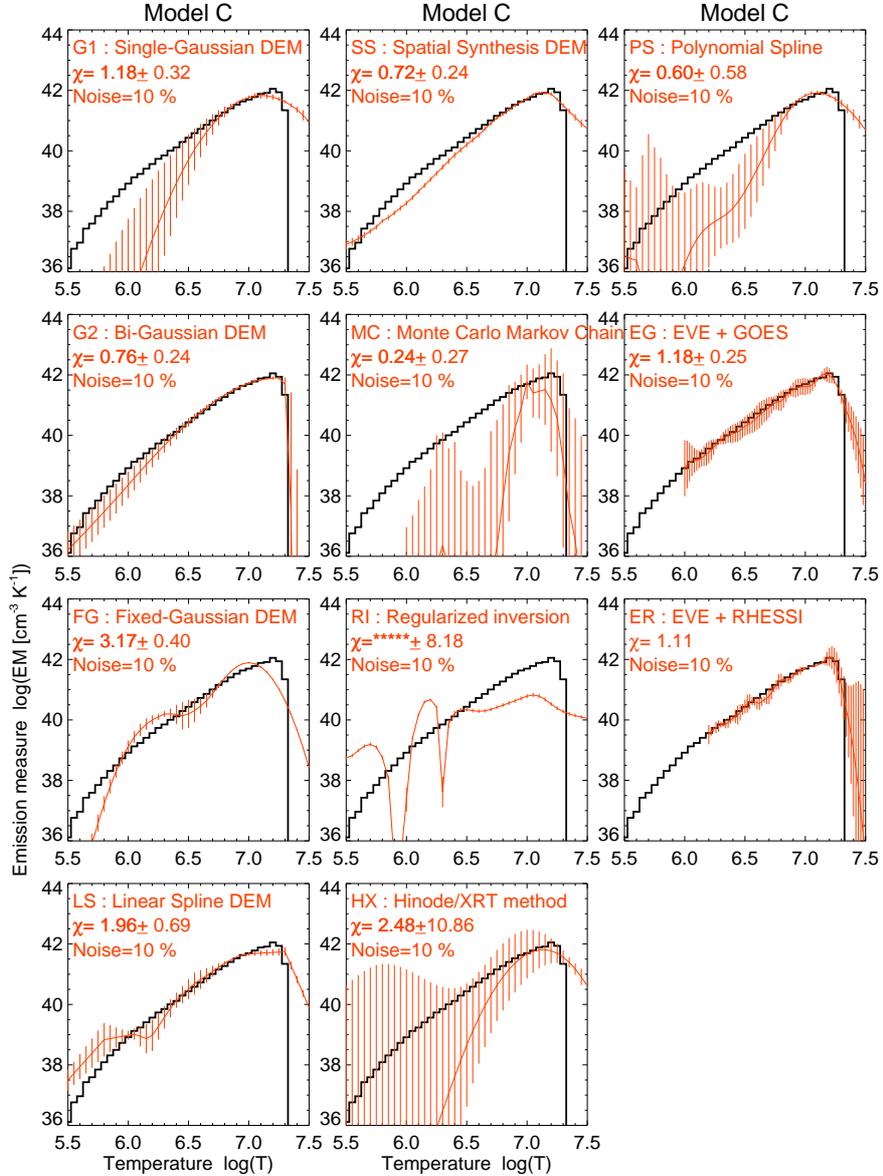}}
\caption{DEM inversions for nanoflare loop model C, using
11 different DEM methods. Otherwise representation similar
to Fig.~5.}
\end{figure}

\begin{figure}
\centerline{\includegraphics[width=1.0\textwidth]{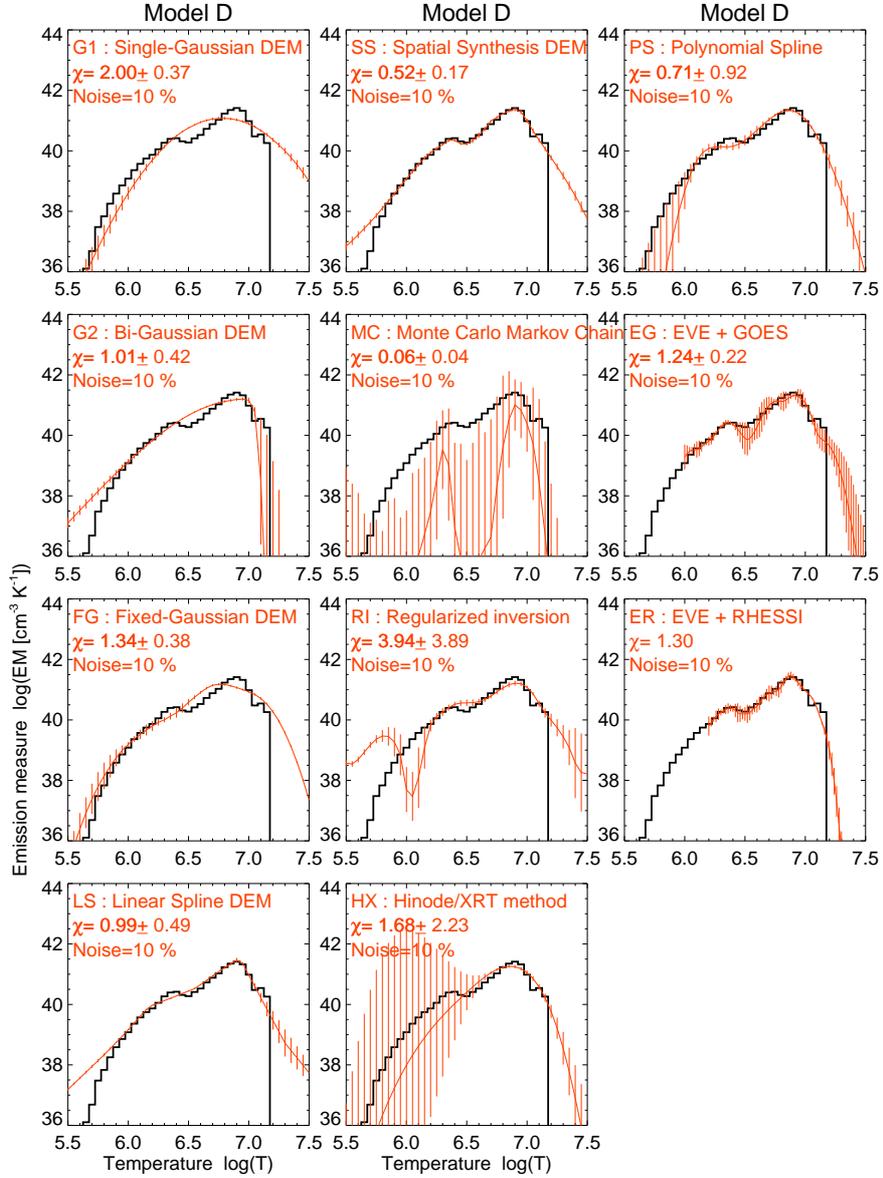}}
\caption{DEM inversions for the dual-temperature arcade 
model D, using 11 different DEM methods. Otherwise 
representation similar to Fig.~5.}
\end{figure}

\begin{figure}
\centerline{\includegraphics[width=1.0\textwidth]{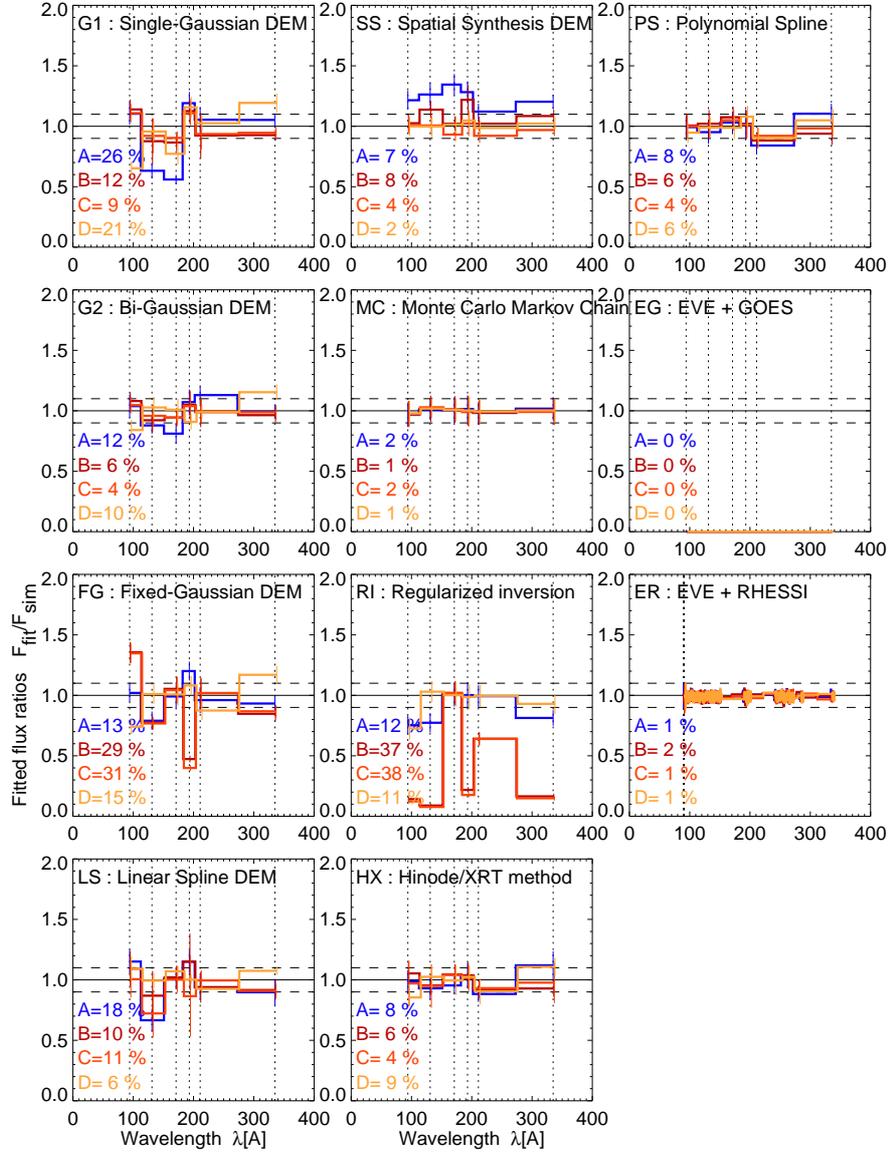}}
\caption{Ratios of fitted fluxes to simulated fluxes for the
6 AIA wavelengths. The error bars were obtained from adding
10\% random noise to the simulated fluxes. The standard
deviation of the fitted to the simulated fluxes are quoted
in percentages. The x-axis represents the
logarithm of the temperature in units of K.}
\end{figure}

\begin{figure}
\centerline{\includegraphics[width=1.0\textwidth]{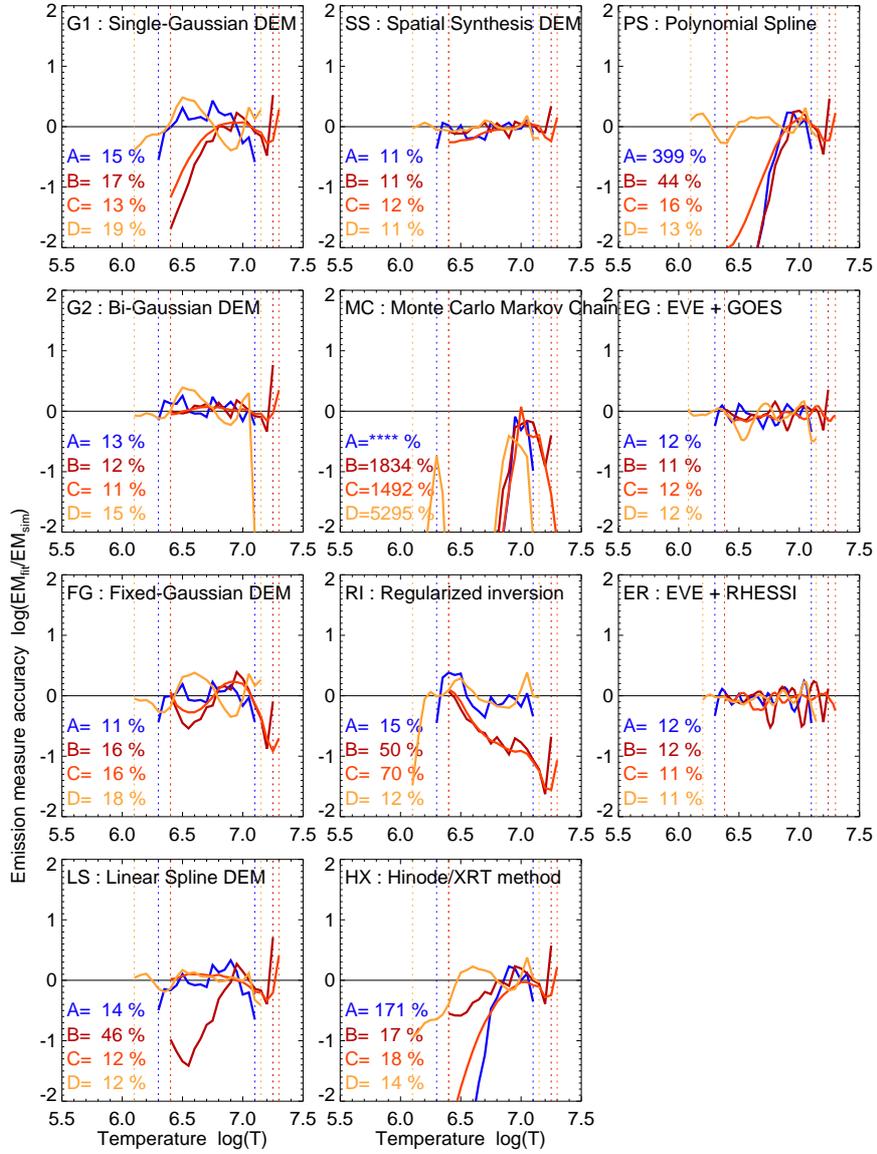}}
\caption{The accuracy of the 11 DEM inversion methods is
shown by the ratio of the best-fit and simulated emission measure,
$\log( EM^{fit}/EM^{sim})$, as a function of the temperature $\log(T)$,
where the color indicates the three simulations A (blue), B (brown),
C (red), and D (orange). The ranges within two orders of magnitude below the
peak emission measure $EM^{sim}$ are indicated with dotted vertical
lines and the average deviations in percentages are listed for this
range. The x-axis represents the temperature in units of K.}
\end{figure}


\begin{table}
\caption{Characteristics of 11 DEM methods applied in this study:
Name of DEM method, DEM parameterization function, degrees of freedom,
optimization algorithm, number of spatial pixels, and instrument of used data.}
\begin{tabular}{llllll}
\hline
Method 	& Parametrization & Degrees 	& Optimization 			& Pixels& Data  \\
	&		  & of		& algorithm			&	&	\\
	&		  & freedom	& 				&	&	\\
\hline
G1	& Single Gaussian & 3		& Least-square minimization	& 1 	& AIA	\\
G2	& Bi-Gaussian 	  & 4		& Least-square minimization	& 1	& AIA	\\
FG	& Fixed Gaussians & 4-6		& Least-square minimization   	& 1	& AIA	\\
LS	& Linear Spline   & 4-6		& Least-square minimization   	& 1	& AIA	\\
SS	& Single Gaussian & 3		& Least-square minimization 	& 16,384& AIA	\\
MC	& N/A		  & 61		& Markov chain Monte Carlo 	& 1 	& AIA	\\
RI	& N/A		  & 61		& Regularized Inversion 	& 1	& AIA	\\
HX	& Spline	  & 4-6		& Least-square minimization	& 1	& AIA	\\
PS	& Spline	  & 4-6		& Least-square minimization	& 1	& AIA	\\
EG	& Fixed Gaussians & 10-20	& Least-square minimization 	& 1	& EVE+ 	\\
  	& 		  & 		&  				& 	& GOES 	\\
ER	& Fixed Gaussians & 10-20	& Least-square minimization	& 1	& EVE+	\\
  	& 		  & 		&  				& 	& RHESSI\\
\hline
\end{tabular}
\end{table}

\section{	DEM INVERSION METHODS AND RESULTS 	}

In the following section, we describe 11 different DEM inversion 
and forward-fitting methods, where each method consists
of a choice of DEM parameterization, degrees of freedom, optimization
algorithm, spatial summing, and data set from different instruments,
as indicated in Table 1 for the cases studied here.
In the following we present also the results that these 11 methods 
produce for the 4 simulated models A-D pictured in Fig.~(1). The 
results are graphically presented in Figs.~(5-10), and quantitatively 
compiled in Tables 2 and 3.

The values of the simulated DEM peak temperature $T_p$, the DEM peak emission 
measure $EM_p$, the total emission measure $EM_{tot}$, the multi-thermal 
energy $E_{th}$, and the $\chi^2$-values of the best fits are specified 
in Table 2 for the 4 models (A-D), while the ratios of the best-fit 
values to the simulated values are compiled in Table 3. 
The ratios of the fitted to the simulated wavelength fluxes are graphically
presented in Fig.~(9), while the (logarithmic) ratios of the simulated to the
inverted DEMs are shown in Fig.~(10).

\subsection{	Single-Gaussian DEM Fit (G1) 		}

One of the most robust choices of a DEM function with a minimum of
free parameters is a single Gaussian (in the logarithm of the
temperature), which has 3 free parameters only and is defined by
the peak emission measure $EM_p$, the DEM peak temperature $T_p$,
and the logarithmic Gaussian temperature width $w$. The DEM
parameter has the cgs-units [cm$^{-5}$ K$^{-1}$],
\begin{equation}
        EM(T) = n_e^2 {dz \over dT} = EM_p \exp{ \left(
        - {[log(T) - log(T_p)]^2 \over 2 w^2 }
        \right)} \ ,
\end{equation}
where the total emission measure $EM = n_e^2 dz = \int EM(T) \ dT$ is
the temperature integral over the Gaussian DEM (in units of [cm$^{-5}$]).
Single-Gaussian DEM fits have been used in many studies with AIA data
(e.g., Aschwanden and Boerner 2011; Aschwanden and Shimizu 2013;
Ryan et al.~2014).

In our DEM forward-fitting code we define logarithmically-spaced 
values for the electron temperature, $log(T_p) = 5.0,...,8.0$ K, 
for the logarithmic Gaussian width $w = 0.01,..., 1$,
and for the peak emission measure values $log(EM_p) = 23, ..., 29$.
The temperature range is identical to the discretized definition of 
the instrumental response function $R(T)$.
Approximate best-fit values are then found simply by a global search in 
this 3-parameter space $[T_p, w, EM_p]$ by calculating the 
minimum of the reduced $\chi^2$-criterion (e.g., Bevington and Robinson 1992),
\begin{equation}
        \chi^2_{red} = \left[
        {1 \over n_{free}}
        \sum_{\lambda=1}^{n_\lambda} {(f_{\lambda}^{fit} - f_{\lambda}^{sim})^2
                \over \sigma_{\lambda}^2} \right]^{1/2} \ ,
\end{equation}
where $f_{\lambda}^{sim}$ are the 6 simulated AIA flux values,
$f_{\lambda}^{fit}$ are the fitted flux values (using the Gaussian EM(T)
defined in Eq.~9), 
\begin{equation}
        f_{\lambda}(T_p, w, EM_p) 
        = \sum_{k=1}^{n_T} EM(T_k, w, EM) \ R_{\lambda}(T_k) 
	  \ \Delta T_k \ ,
\end{equation}
$\sigma_\lambda$ are the estimated
uncertainties, $n_{free}=n_{\lambda}-n_{par}$ is the number
of degrees of freedom, which is $n_{free}=3$ for $n_{\lambda}=6$,
the number of AIA wavelength filters, and $n_{par}=3$, the number of
model parameters. The uncertainties $\sigma_{\lambda}$, which are
dominated by the inaccuracies in the knowledge of the AIA instrumental response
function, are estimated to be of the order of 10\% of the 
observed fluxes (Boerner et al.~2014). We neglect 
the Poisson noise of the photon statistics, which is on the order
of $\sqrt{(N)}/N \approx 10^{-4}-10^{-3}$ for AIA count rates
of $N \approx 10^6-10^8$ photons/s (Boerner et al.~2014; O'Dwyer
et al.~2010). Consequently, the estimate of the uncertainty is 
\begin{equation}
	\sigma_{\lambda} \approx q_{noise} \ f_{\lambda}^{sim} \ ,
\end{equation}
with $q_{noise} \approx 0.1$. We generate simulated fluxes by adding
10\% random noise to the noise-free data, i.e., 
$f_{\lambda}^{ran} = (1 + \rho_\lambda \ q_{noise}) \ f_{\lambda}^{sim}$,
where random values $\rho_\lambda$ are drawn from a normal distribution with a 
standard deviation of unity.  From a global search of the minimum in the 
3-parameter space $[T_p, w, EM_p]$ we obtain an approximate solution for 
the DEM, which we use as initial guess for a refined optimization using 
the Powell $\chi^2$-minimization method (Press et al.~1986).

\medskip \underbar{\bf Results G1:}
We turn now to the results of our DEM fits.
The best fits to the simulated fluxes $f_{\lambda}^{sim}$ (Eq.~6) of a
single-Gaussian DEM function are shown in Figures 5-8 (top left panel), 
which fit the simulated DEM function most accurately near the peak of
the DEM, but yield too low DEM values at the low-temperature side and 
too high DEM values at the high-temperature side.  We see that the 
model-predicted fitted flux values differ by 9\%-26\% with respect 
to the simulated input flux values for the four cases A-D (Fig.~9), 
and the resulting DEM functions differs by 13\%-19\% (Fig.~10). 
Obviously, a single-Gaussian DEM function cannot fit a highly asymmetric
DEM well, but is likely to do better for a near-symmetric
single-peaked DEM distribution function. This is also consistent with the
goodness-of-fit criterion, which yields values significantly above unity
for all 4 cases A-D (Fig.~5-8), indicating that the G1 model does not fit
the data as well as possible within the estimated flux uncertainties. 

\subsection{	Bi-Gaussian DEM Fit (G2)		}

In order to accommodate asymmetric DEM functions, we introduce the
bi-Gauss\-ian DEM function (G2), which is composed of two semi-Gaussian
functions with two different temperature half widths $w_1$ and $w_2$ 
on the left and right side, 
\begin{equation}
        EM(T) = EM_p \exp{ \left(
        - {[log(T) - log(T_{p})]^2 \over 2 \ w^2 }
        \right) } \quad 
	\left\{ \begin{array}{ll}
		w = w_1\ &{\rm for}\ T \le T_p \\
		w = w_2\ &{\rm for}\ T \ge T_p \\
		\end{array}
	\right.
   	\ ,
\end{equation}
with a total of four variables ($n_{par}$), i.e., $EM_p, T_p, 
w_1, w_2$. Since the AIA instrument has $n_{par}=6$ coronal
wavelength filters, there are two degrees of freedom, 
$n_{free}=n_{\lambda}-n_{par} = 6 - 4 = 2$, for this model.

We choose equidistant values in the logarithmic ranges of 
$log(EM_p)=23,...,$ 29 for the emission measure (cm$^{-5}$ K$^{-1}$), 
$log(T_p)=5.5,...,7.5$ for the peak temperatures (K), and 
$w=0.01,...,10$ for the Gaussian widths. As with the 
single-Gaussian method, G1, we find first an approximate
DEM solution from an absolute search in the discretized three-dimensional 
parameter space, which is then used as initial guess for subsequent 
optimization with the Powell $\chi^2$-minimization method (Press et al.~1986).

\medskip \underbar{\bf Results G2:}
The best fits are shown in Figs.~(5)-(10) (second left panels). All of 
the fits for the four models A-D represent the peak of the asymmetric DEM 
function better than the (symmetric) single-Gaussian fits from G1. 
The low-temperature side of the DEM function can be fitted with a 
large semi-Gaussian width of $w1 \approx 1.5$
and the high-temperature side with a very small semi-Gaussian width of
$w_2 \approx 0.01$ to mimic the steep drop-off. 
The goodness-of-fit criterion varies in the range of $\chi^2 \approx 0.76-1.33$
for the four cases A-D (Figs.~5-8; second left frames),
the fitted flux values differ by 4\%-12\% with respect to the simulated 
flux values (Fig.~9), and the resulting DEM functions differ by 11\%-15\% 
(Fig.~10).  The actual uncertainty of the AIA response function is 
comparable, i.e., $\sim 10\%$ (Boerner et al.~2014), which we emulate 
here by generating 30 data sets of fluxes with 10\% noise added.
The bi-Gaussian DEM function turns out to be
a good choice for the asymmetric single-peaked DEM models A-C used here,
but is by nature less adequate to fit multi-peaked DEMs (such as model D). 

\subsection{	Multiple Fixed-Gaussian DEM Fit (FG)	}

In principle, one can define a more versatile DEM function by a superposition
of an arbitrary number of Gaussian functions (Eq.~9, $n > 2$), but the number
of free parameters will increase by $n_{free} = 3n$, which becomes prohibitive
for practical purposes of numerical fitting. One way to keep the number of
free parameters within a reasonable limit is to fix the peak temperatures
$T_{p,i}$ at logarithmically equi-spaced values (with a step of $\Delta log(T)$)
and to fix the Gaussian widths to a value that corresponds to about half the
separation step, i.e., $w = \Delta log(T)/2$, so that
only the emission measure values $EM_{p,i}$ have to be optimized. Such
a multiple fixed-Gaussian method (FG) has been applied in several studies using
AIA or EVE data (e.g., Aschwanden and Boerner 2011; Warren et al.~2013; 
Caspi et al.~2014b; Warren 2014). In this section, we apply this method 
to the AIA data alone; the application to combined EUV and X-ray data, 
using EVE and GOES/XRS and/or RHESSI together, is discussed in later sections. 

\medskip \underbar{\bf Results FG:}
We show the application of such a multiple fixed-Gaussian DEM method to the
three models A, B, and C in Figs.~(5-10) (third left panel), using six 
fixed Gaussians centered at the temperatures $log(T)$=6.00, 6.25, ..., 7.25 
with a fixed width of $w=0.13$. The best fits show a double-humped DEM 
function, which seem to depend
somewhat on the choice of the fixed temperatures $T_{p,i}$. In any case, the
DEM function with fixed Gaussians appears to fit the examples here with less
accuracy than the bi-Gaussian function.  
The goodness-of-fit criterion varies in the range of $\chi^2 \approx 1.4-3.2$
for the four cases A-D (Figs.~(5-8), third left frames),
the fitted flux values differ by 13\%-31\% with respect to the simulated 
flux values (Fig.~9), and the resulting DEM functions differ by 11\%-18\% 
(Fig.~10). 
We conclude that DEM functions with multiple fixed Gaussians can provide good
fits when the fixed temperatures agree with the peaks of the DEM, which 
requires some fiddling of the fixed peak temperature values.
However, we note that the application of this method with EVE+GOES and 
EVE+RHESSI data yields superior fits (see below), most likely because 
of the significantly greater EUV spectral data available from EVE, 
combined with the strong constraints on the high-temperature tail 
provided by the X-ray data, which also allow a greater number of 
fixed Gaussian components to be used in the model.

\subsection{	Linear Spline DEM Fit (LS)	}

Spline functions are defined by fixed points $x_i$ on the x-axis, which 
have some particular values $y_i = y(x_i)$ on the y-axis, and are linearly 
interpolated in between. We choose 5 temperature spline points  
$T_k, k=1,...,5$ on the logarithmic x-axis covering the range of 
$T_k =10^{5.8},...,10^{7.3}$ K, and consider the corresponding 5 spline 
values $EM_k, k=1,...,5$ as free parameters to be fitted, which is very 
similar to the multiple fixed-Gaussian fitting method. The spline points 
have a fixed step of $\Delta log(T_k) = 0.375$.  We find an initial guess 
with an absolute minimum search in the five-dimensional 
$\chi^2$-space, followed by subsequent Powell optimization. 

\medskip \underbar{\bf Results LS:}
We show the application of such a linear spline DEM fitting method to the
four models A-D (Figs.~5-8), 
The good\-ness-of-fit criterion varies in the range of $\chi^2 \approx 1.5-2.0$
for the 4 cases (Figs.~(5-8),
the fitted flux values differ by 6\%-18\% with respect to the simulated 
flux values (Fig.~9), and the resulting DEM functions differ by 12\%-46\% (Fig.~10). 
In conclusion, the linear spline function fits the simulated DEMs with similar
accuracy as the multiple fixed-Gaussian method, but not as good as the bi-Gaussian
function in these particular simulations. However, for the general case with multiple 
DEM peaks, the linear spline function is expected to be more flexible than a 
single-Gaussian or bi-Gaussian function.

\subsection{	Spatial Synthesis DEM Method (SS)		}

The DEM methods used so far all use the total spatially-integrated flux from an observed
area that corresponds to the flare or active region size, and thus
contains many locations with possibly quite different temperature structures.
Applying a DEM inversion to a single pixel of an image, we are more
likely to disentangle the complex multi-temperature structures into
simpler components, in some cases as simple as a near-isothermal
segment of a resolved monolithic loop, in which case a Gaussian fit
becomes more adequate. 

A sensible approach to the temperature discrimination 
problem is to subdivide the image area of a flare into macro-pixels
or even single pixels, and then to perform a forward-fit of a
single-Gaussian DEM function in each spatial location separately,
while the total DEM distribution function of the entire flare area
can then be constructed by summing all DEM functions from each
spatial location, which we call the ``Spatial Synthesis DEM'' method.
This way, the Gaussian approximation of a DEM
function is applied locally only, but can
adjust to different peak emission measures and temperatures at each
spatial location. Such a macro-pixel algorithm for automated
temperature and emission measure analysis has been developed
for  AIA wavelength-filter images in Aschwanden et al.~(2013),
and a SSW/IDL code is available online
({\sl http://www.lmsal.com/$\sim$aschwand/ software/aia/aia$\_$dem.html}).
The flux $F_{\lambda}(x,y,t)$ is then measured in each macro-pixel
location $[x,y]$ and the fitted DEM functions
are defined at each location $[x,y]$ separately,
\begin{equation}
        EM(x,y; T) = EM_p(x,y) \exp{ \left(
        - {[log(T) - log(T_p[x,y])]^2 \over 2\ w^2[x,y] }
        \right)} \ ,
\end{equation}
and are forward-fitted to the observed fluxes $F_{\lambda}(x,y)$
at each location $(x,y)$ separately, after background subtraction 
of $B_{\lambda}(x_i, y_j)$ (which is set to zero in our simulated
cases here), 
\begin{equation}
        F_{\lambda}(x_i,y_j) - B_{\lambda}(x_i,y_j) =
        \sum_{k=1}^{n_T} EM(x_i ,y_j; T_k) R_{\lambda}(T_k) \ .
\end{equation}
The synthesized differential emission measure distribution $EM(T)$
can then be obtained by summing up all local DEM distribution functions
$EM(T; x,y,t)$ (in units of cm$^{-3}$ K$^{-1}$),
\begin{equation}
        EM(T) = \sum_i \sum_j EM(x_i, y_j; T_k)\ dx_i dy_j \ ,
\end{equation}
and the total emission measure of a flaring region is then obtained
by integration over the temperature range (in units of cm$^{-3}$),
\begin{equation}
        EM   = \sum_k EM_k(T_k) \ \Delta T_k \ .
\end{equation}
Note that the synthesized DEM function $EM(T)$ (Eq.~16) generally
deviates from a Gaussian shape, because it is constructed from the
summation of many Gaussian DEMs from each macro-pixel location with
different emission measure peaks $EM_p(x_i, y_j)$, peak temperatures
$T_p(x_i, y_j)$, and thermal widths $w(x_i, y_j)$. This
synthesized DEM function can be arbitrarily complex and accommodate
a different Gaussian DEM function in every spatial location $(x_i, y_j)$.

We developed a numerical code for this spatial-synthesis (SS) DEM method.
We extract a field-of-view of $512 \times 512$ pixels from the observed 
AIA images in 6 coronal wavelengths. We rebin them by 
macro-pixels with a binsize of 4 full-resolution pixels,
which forms a grid of 128$\times$128 macro-pixels $[x_i, y_j]$.
The code performs then $128^2=16,384$ single-Gaussian DEM fits per 
wavelength set, and adds up the partial DEMs of each macro-pixel to 
a total field-of-view DEM function. 

\medskip \underbar{\bf Results SS:}
The results of the DEM inversion of our model A-D are shown 
in Figs.~(5-10) (top middle panels). 
It appears that the detailed
shape of the DEM function is well-fitted with the spatial synthesis 
DEM method in all 4 cases A-D.
The goodness-of-fit criterion varies in the range of $\chi^2 \approx 0.5-2.5$,
the fitted flux values differ by 2\%--8\% with respect to the simulated 
flux values (Fig.~9), and the resulting DEM functions 
differ by 11\%-12\% (Fig.~10). 
Given the estimated uncertainties of the response function in the order of
$\sim 10\%$ (Boerner et al.~2014), the spatial synthesis DEM method 
appears to provide a most adequate and flexible parameterization that 
fits the DEM within the uncertainties of the response functions. Tests
of over 6400 DEM reconstructions during about 400 flares
demonstrate that arbitrarily complex DEMs can be adequately fitted with 
the spatial synthesis method (Aschwanden et al.~2015), 
regardless whether the DEMs are asymmetric or multi-peaked, which 
generally cannot be achieved with the previously discussed DEM methods.
Of course, if AIA data are used alone, the temperature range of
reliable DEM reconstruction is bound to $\log(T) \approx 5.8-7.3$.

\subsection{	Monte-Carlo Markov Chain Method (MC)	}

The {\sl Monte Carlo Markov chain (MC)} method is a forward-modeling
DEM method that does not assume a particular functional form for the
DEM distribution function (Kashyap and Drake 1998), but applies some
smoothness criteria which are locally variable and based on the
properties of the temperature responses and emissivities for the
input data, instead of being arbitrarily determined a priori
(Testa et al.~2012a). The MC method is considered to produce robust
and unbiased results, because it does not impose a pre-determined
or arbitrarily selected functional form for the solution, and
because it provides also an estimate of the uncertainties of the
DEM function. The MC method has been applied to solar DEM modeling
of {\sl Solar Extreme Ultraviolet Rocket Telescope and Spectrograph
(SERTS)} observations (Kashyap and Drake 1998), to SoHO and Hinode
data (Landi et al. 2012), to Hinode/EIS, XRT, and SDO/AIA data
(Hannah and Kontar 2012; Testa et al.~2012a), as well as to
artificial data simulated with an MHD code (Testa et al.~2012a).  
The latter application provides a validity test, since the exact
DEM solution is known from the simulated data, which revealed that
the differences between the simulated and forward-fitted data were
larger than predicted by the MC method (Testa et al.~2012a). 

\medskip \underbar{\bf Results MC:}
The application of the MC method to our 4 simulated models A-D 
reveals two striking properties. First, the fluxes are fitted
extremely well, with an accuracy of 1\%-2\% (Fig.~9), which is
the highest accuracy obtained in our benchmark test. Secondly,
the DEM function appears to be fitted poorly, underestimating
the simulated DEM function at all temperatures except near the
peak of the DEM (Figs.~5-8). The goodness-of-fit values are in
the range of $\chi^2=0.06-0.25$ (Fig.~5-8), which indicates that
the MC method over-fits the data. The other methods described here
all apply a smoothness constraint on their result in form of a
parameterization of the DEM function; this has the effect of
limiting the accuracy in the recovery of the input fluxes
in the presence of noise or non-smooth input DEMs, but it also
tends to produce solutions that are more robust to small signal
fluctuations. The MC DEM results are highly sensitive to noise
in the observations; however, the algorithm provides an estimated
uncertainty, which in general is close to the difference between
the simulated and fitted DEM, while the average uncertainty
(obtained here from 30 iterations with 10\% added noise) is mostly
below. 

\subsection{	Regularized Inversion Method (RI)	}

Similar to the MC inversion algorithm, the {\sl Regularized
Inversion (RI)} method, developed by Hannah and Kontar (2012), 
introduces an additional ``smoothness'' to constrain the
amplification of the uncertainties, allowing a stable inversion
to recover the DEM solution. The RI spectral inversion method
was first applied to RHESSI spectra (e.g., Piana et al.~2003;
Massone et al.~2004; Brown et al.~2006; Prato et al.~2006),
and subsequently to SDO/AIA and Hinode/EIS data (Hannah
and Kontar 2012). The RI method was tested with Gaussian,
multi-Gaussian, and CHIANTI DEM models. The uncertainty in
finding the peak temperature was found to be of order
$\Delta log(T) \approx 0.1-0.5$, depending on the noise level.
For Hinode/EIS data, the DEM is broadened by an uncertainty 
of $\sim 20\%$. For active region DEMs observed with
EIS and XRT, the RI method was found to retrieve
a similar DEM as the MC method (Hannah and Kontar 2012).

\medskip \underbar{\bf Results RI:}
In our application of the regularized inversion method we
found a passable DEM fit for the models A and D, while
the method failed and did not converge for model B and C (Fig.~5-8).
The fitted flux ratios have an accuracy of $\sim 11\%-12\%$
for model A and D (Fig.~9), but a much poorer accuracy
of $\sim 37\%-38\%$ for model B and C (Fig.~9). Also
the ratio of DEM values is acceptable for model A and D
($\sim 12\%-15\%$), while it is unacceptable for model
B and D with $\sim 50\%-70\%$ (Fig.~10). 
Comparing among all 11 tested DEM algorithms, the
regularized inversion method shows the poorest
performance for the particular 4 models tested here.
The RI method appears to perform best for model D. 
In the other cases, the DEM is strongly peaked at temperatures 
above the regime where AIA is most sensitive and best able to
discriminate ($\log(T)=6.0-6.5$). As a result, most of the
signal in the AIA channels is not due to material at or near the
temperature of the channel's peak sensitivity. This poses a 
difficult challenge for inversion methods.

\subsection{	Hinode/XRT Method (HX)	                }

The {\sl Hinode/XRT} algorithm is a DEM forward-fitting
method that uses the functional form of a spline function
(i.e., a small number of spline points $EM(T)$ interpolated
by some polynomial function). DEM fitting with spline functions
has been explored using Skylab/XUV spectrograph data 
(Monsignori-Fossi and Landini 1992), SERTS data (Brosius et 
al.~1996), SOHO/CDS and UVCS data (Parenti et al.~2000), and more 
recently using SDO/AIA and Hinode/XRT data (Weber et al.~2004;
Golub et al.~2007). The spline points are adjusted iteratively 
as part of the minimization routine, unlike the fixed-spline 
points used in the LS method. The forward-fitting uses the IDL 
routine {\sl mpfit.pro} and is implemented in the IDL routine 
{\sl xrt$\_$dem$\_$iterative2.pro}.

\medskip \underbar{\bf Results HX:}
Our application of the HX method to the four models A-D shows 
that the DEM algorithm always recovers the peak of the DEM 
function well, but retrieves the DEM function to a much less 
accurate degree in the low-temperature regime $\log(T_e)
\approx 5.5-6.5$. Nevertheless, the flux ratios are 
retrieved very accurately (with $\approx 4\%-9\%$; Fig. 9), 
while the DEM values are underestimated for model A, 
but within the error estimates for the models B-D.

\subsection{	Polynomial Spline Method (PS)	        }

This is an alternative polynomial spline algorithm developed 
primarily for SDO /AIA data. Its basic approach is quite 
similar to the HX method; it was developed independently, 
and thus the details of the implementation are slightly 
different (for example, it uses a variable number of spline 
points, instead of setting the number to one less than the 
number of observations). This routine was used in the initial 
calibration of SDO/AIA (Boerner et al.~2012) and later in 
photometric and thermal cross-calibrations of AIA with 
Hinode/ESI and Hinode/XRT (Boerner et al.~2014). The 
difference between this method and the HX method can be 
thought of as providing a feel for the potential significance 
of low-level differences in a code on the results.

\medskip \underbar{\bf Results PS:}
The application of the PS method to our four models A-D
shows acceptable fits near the DEM peak at 
$log(T_e) \approx 7.0$ (Figs.~5-8). The fitted/simulated
flux ratios are obtained with a high accuracy of
$\sim 4\%-8\%$ (Fig.~9), and the DEM ratios are
accurate at high temperatures (Fig.~10), but underestimate
the DEM somewhat at lower temperatures.

\subsection{	EVE and GOES Fitting Method (EG)	}

The previously described tests all (with the exception of HX) 
use the AIA response functions (Fig.~3) to retrieve the DEM
distribution function, which works generally well in the 
temperature range of $log(T_e) \approx
5.5-7.0$, but is insufficiently covered at high flare 
temperatures at $log(T_e) \approx 7.0-7.5$.
The high-temperature regime up to $log(T_e) \approx 7.5$, 
however, is well-covered by the combination of the SDO/EVE 
and GOES/XRS instruments (the high-temperature tail above 
$\log(T) \approx 7.5$ is typically poorly observed as its 
low emission measure means it contributes relatively little 
to the overall GOES/XRS signal). As the
temperature range in Fig.~4 shows, EVE has high-temperature 
lines with $log(T_e) \gapprox 7.1$ in
the wavelength regime of $\lambda \approx 90-270$ \ang .

A DEM reconstruction method using {\sl EVE and GOES (EG)} 
data has been developed and described in
Warren et al.~(2013) and Warren (2014).  A DEM function 
composed of $n_G=18$ Gaussian at fixed
temperatures is chosen (similar to the FG method 
described in Section 4.3), which is then
convolved with the atomic (CHIANTI) emissivities and 
yields a modeled EVE spectrum, as well as
GOES fluxes for the 0.5-4 \ang\ and 1-8 \ang\ channels. 
The magnitudes of the emission measures,
$EM(T_i), i=1,...,n_G$, are then varied and optimized 
until the EVE spectrum and GOES fluxes are
satisfactorily reproduced.  A $\chi^2$ goodness-of-fit 
criterion is calculated from the fits to
the EVE and GOES fluxes. As Fig.~4 shows, the EVE fluxes 
consists of both continuum and line emission, which are 
both calculated from the CHIANTI emissivities in the 
modeled EVE irradiance spectrum.

\medskip \underbar{\bf Results EG:} The application of the 
EG method to our four models A-D shows
that the simulated DEM functions are retrieved very 
well in all 4 cases, with $\chi^2$-values
in the range of $\chi^2=1.1-1.3$ (Figs.~5-8). Also the 
estimated errors of the recovered DEM
functions are comparable with the differences of 
fitted/simulated values (Fig.~10). The ratio of
the fitted/simulated emission measure values are in the 
range of $\sim 11\%-12\%$ (Fig.~10),
evaluated from 30 different runs with 10\% added random noise.
Adding 5\% noise instead of 10\% did not make any difference.

\subsection{	EVE and RHESSI DEM Fitting Method (ER)	}

A similar combination as the EG method, is the combination
of {\sl SDO/EVE and RHESSI (ER)} data. EVE is sensitive in
the temperature range of $T_e \approx 2-25$ MK, and RHESSI
is sensitive to thermal emission at $T_e \gapprox 10$ MK. 
While the EG method has limited accuracy above $\sim$30 MK, 
RHESSI is increasingly sensitive to higher temperatures and, 
as a spectrometer, can also resolve broad DEM structures 
more accurately than the two-channel GOES/XRS data. The ER 
method is therefore applicable to the broadest temperature 
range, from $\sim$2 MK up to the hottest temperatures existing 
(typically, $\sim$50 MK in the most intense flares; Caspi et al. 2014a).
The ER method has been tested and applied to two solar flares
in Caspi et al.~(2014b). Similar to the EG method, the DEM
distribution function is composed of multiple (typically 10-20) 
Gaussians at fixed temperatures (like in the FG method), 
which are used to generate predicted photon fluxes that are then
convolved with the RHESSI detector response function and
the CHIANTI atomic emissivities to obtain RHESSI X-ray and 
EVE MEGS-A EUV irradiance spectra. The emission measures 
of the Gaussian components are then varied to minimize the 
$\chi^2$, as in the FG and EG methods.

\medskip \underbar{\bf Results ER:}
Applying the ER method to our four DEM models A-D,
we find acceptable fits for all four cases, with 
goodness-of-fit criteria of $\chi^2=1.1-2.0$
(Figs.~5-8). The EVE and RHESSI fluxes are very well
matched (within an accuracy of $\sim 1\%$) (Fig.~9).
Also the ratio of the fitted/simulated DEM values is
well matched ($10\%-12\%$; Fig.~10). The uncertainties
of the fitted DEM values is estimated from 30 runs with
10\% added random noise and is comparable with the actual
difference between fitted and simulated DEM values
(Fig.~5-8, 10). Adding 5\% noise instead of 10\%
did not make any difference.


\bigskip
\begin{table}
\caption{The values of the simulated DEM Peak temperature $T_p$ (MK), 
DEM peak emission measures $EM_p$ (cm$^{-3}$ K$^{-1}$), total emission 
measures $EM_{tot}$ (cm$^{-3}$), and total multi-thermal energies 
$E_{th}$ (erg) of the simulated models A, B, C, and D.}
\begin{tabular}{lllll}
\hline
DEM 	  	    & DEM 		& DEM peak	& Total		& Multi-  \\
method		    & peak 		& emission	& emission 	& thermal \\
		    & temperature 	& measure	& measure	& energy  \\
  		    & log($T_p$) 	& log($EM_p$)   & log($EM_{tot}$)&log($E_{th}$) \\
\hline
Model A :  &  7.100 & 41.993 & 48.700 & 29.901  \\
Model B :  &  7.200 & 43.038 & 49.779 & 30.568  \\
Model C :  &  7.200 & 42.053 & 49.003 & 30.242  \\
Model D :  &  6.900 & 41.416 & 48.103 & 29.600  \\
\hline
\end{tabular}
\end{table}


\begin{table}
\caption{The ratios of the fitted to the simulated values of 
the DEM Peak temperature, $q_T=T_p^{fit}/t_p^{sim}$, 
the DEM peak emission measures $q_{EMp}=EM_p^{fit}/EM_p^{sim}$ (cm$^{-3}$ K$^{-1}$), 
the total emission measures    $q_{EMt}=EM_t^{fit}/EM_t^{sim}$ (cm$^{-3}$), 
the multi-thermal energy       $q_{Eth}=E_{th}^{fit}/E_{th}^{sim}$ (erg),
the goodness-of-fit $\chi^2$, and 
the fluxes $q_f = f_\lambda^{fit}/f_\lambda^{sim}$, 
tabulated for 4 DEM models (A,B,C,D) and 11 DEM reconstruction 
methods.}
\begin{tabular}{llrrrrrr}
\hline
DEM	& DEM 	& DEM 	& DEM           & Total		& Multi-  & Goodness & Flux  \\
model	& method& peak 	& emission	& emission 	& thermal & of fit   & ratio \\
   	&       & temp. & measure	& measure	& energy  &          &       \\
      	&       & ratio	& ratio		& ratio         & ratio   &          &       \\
  	&       & $q_T$ & $q_{EMp}$     & $q_{EMt}$     &$q_{Eth}$& $\chi^2$ & $q_F$ \\
\hline
A & G1 &    0.56 &   0.52 &    1.00 &   1.61 &    2.65$\pm$   0.37 &    0.93$\pm$   0.27\\
A & G2 &    1.00 &   0.59 &    0.89 &   0.96 &    1.33$\pm$   0.33 &    0.99$\pm$   0.12\\
A & FG &    0.79 &   0.53 &    0.99 &   1.50 &    1.36$\pm$   0.37 &    0.98$\pm$   0.13\\
A & LS &    0.63 &   0.67 &    0.95 &   1.54 &    1.83$\pm$   0.32 &    0.97$\pm$   0.18\\
A & SS &    1.00 &   0.80 &    0.99 &   1.32 &    2.53$\pm$   0.21 &    1.24$\pm$   0.08\\
A & MC &    0.79 &   0.38 &    0.30 &   0.45 &    0.22$\pm$   0.26 &    1.00$\pm$   0.02\\
A & RI &    0.89 &   0.56 &    0.87 &   1.34 &    7.94$\pm$   4.65 &    0.89$\pm$   0.12\\
A & HX &    0.79 &   0.78 &    0.97 &   1.25 &    1.64$\pm$   1.11 &    0.98$\pm$   0.08\\
A & PS &    0.79 &   0.84 &    0.97 &   1.28 &    1.28$\pm$   0.94 &    0.99$\pm$   0.09\\
A & EG &    0.87 &   0.67 &    0.85 &   1.76 &    1.37$\pm$   0.56 &                0.09\\
A & ER &    0.87 &   0.98 &    0.83 &   1.54 &    1.26             &                0.02\\

\hline
B & G1 &    0.79 &   0.40 &    1.05 &   1.56 &    1.55$\pm$   0.32 &    0.98$\pm$   0.12\\
B & G2 &    1.12 &   0.47 &    0.94 &   1.04 &    0.95$\pm$   0.28 &    0.99$\pm$   0.06\\
B & FG &    0.63 &   0.48 &    0.71 &   0.96 &    2.96$\pm$   0.39 &    0.92$\pm$   0.30\\
B & LS &    1.12 &   0.41 &    1.10 &   1.49 &    1.50$\pm$   0.44 &    1.00$\pm$   0.11\\
B & SS &    1.00 &   0.80 &    0.99 &   1.25 &    1.25$\pm$   0.17 &    1.08$\pm$   0.08\\
B & MC &    0.79 &   0.32 &    0.32 &   0.53 &    0.25$\pm$   0.37 &    1.00$\pm$   0.02\\
B & RI &    0.71 &   0.05 &    0.11 &   0.50 &  149.76$\pm$   9.36 &    0.38$\pm$   0.37\\
B & HX &    0.79 &   0.50 &    1.09 &   1.50 &    0.97$\pm$   0.64 &    0.99$\pm$   0.06\\
B & PS &    0.71 &   0.55 &    1.01 &   1.39 &    1.09$\pm$   0.78 &    0.99$\pm$   0.07\\
B & EG &    0.91 &   0.58 &    0.90 &   1.75 &    1.19$\pm$   0.27 &                0.07\\
B & ER &    0.83 &   0.90 &    0.86 &   1.55 &    1.11             &                0.02\\
\hline
C & G1 &    0.79 &   0.60 &    1.06 &   1.40 &    1.18$\pm$   0.32 &    0.99$\pm$   0.09\\
C & G2 &    1.12 &   0.71 &    0.98 &   1.02 &    0.76$\pm$   0.24 &    0.99$\pm$   0.04\\
C & FG &    0.63 &   0.69 &    0.63 &   0.79 &    3.17$\pm$   0.40 &    0.91$\pm$   0.32\\
C & LS &    1.26 &   0.50 &    0.88 &   1.19 &    1.96$\pm$   0.69 &    0.92$\pm$   0.11\\
C & SS &    0.89 &   0.74 &    1.01 &   1.30 &    0.72$\pm$   0.24 &    0.98$\pm$   0.04\\
C & MC &    0.63 &   0.52 &    0.25 &   0.40 &    0.24$\pm$   0.27 &    1.00$\pm$   0.02\\
C & RI &    0.71 &   0.06 &    0.09 &   0.39 &  157.22$\pm$   8.18 &    0.36$\pm$   0.38\\
C & HX &    0.89 &   0.57 &    0.87 &   1.23 &    2.48$\pm$  10.86 &    0.98$\pm$   0.04\\
C & PS &    0.79 &   0.74 &    1.04 &   1.32 &    0.60$\pm$   0.58 &    0.99$\pm$   0.04\\
C & EG &    0.91 &   0.93 &    0.86 &   1.63 &    1.18$\pm$   0.25 &                0.04\\
C & ER &    1.00 &   1.05 &    0.90 &   1.53 &    1.11             &                0.02\\
\hline
D & G1 &    0.79 &   0.46 &    0.93 &   1.55 &    2.00$\pm$   0.37 &    0.96$\pm$   0.21\\
D & G2 &    1.12 &   0.61 &    0.88 &   0.85 &    1.01$\pm$   0.42 &    0.99$\pm$   0.11\\
D & FG &    0.71 &   0.58 &    0.96 &   1.35 &    1.34$\pm$   0.38 &    0.98$\pm$   0.15\\
D & LS &    1.00 &   1.13 &    1.03 &   1.12 &    0.99$\pm$   0.49 &    1.03$\pm$   0.06\\
D & SS &    1.00 &   0.92 &    0.99 &   1.20 &    0.52$\pm$   0.17 &    1.01$\pm$   0.02\\
D & MC &    1.00 &   0.39 &    0.19 &   0.32 &    0.06$\pm$   0.04 &    1.00$\pm$   0.01\\
D & RI &    1.00 &   0.63 &    0.89 &   1.25 &    3.94$\pm$   3.89 &    0.94$\pm$   0.11\\
D & HX &    0.89 &   0.70 &    0.98 &   1.14 &    1.68$\pm$   2.23 &    0.99$\pm$   0.09\\
D & PS &    0.89 &   0.85 &    1.03 &   1.14 &    0.71$\pm$   0.92 &    0.99$\pm$   0.06\\
D & EG &    1.05 &   0.81 &    0.85 &   1.59 &    1.24$\pm$   0.22 &    0.99$\pm$   0.06\\
D & ER &    0.95 &   1.14 &    0.90 &   1.50 &    1.30             &                0.02\\
\hline
\end{tabular}
\end{table}

\section{	Summary of Results 			}

This study analyzes a set of 11 separate  
DEM methods, using SDO/AIA, as well as SDO/EVE, RHESSI, and GOES data. 
We provide here an objective comparison between the 11 methods by evaluating
their fidelity in matching the EM-weighted temperatures $T_w$, the peak emission
measure $EM_p$, the total emission measure $EM_t$, the multi-thermal energy
$E_{th}$, the goodness-of-fit criterion $\chi^2$, and the flux ratios $q_f$ of the
fitted to simulated data. The results are also
shown in Figs.~(5-10) and Table 3. We summarize here the results of fitting
these test parameters:

\begin{enumerate}
\item{\underbar{\bf EM-weighted temperature $T_w$:} 
By averaging the 11 DEM fits in all 4 cases (i.e., the 44 values in the
third column of Table 3), we find that the emission measure-weighted temperatures
have been determined with an accuracy of $T_w^{fit}/T_w^{sim}$ $=0.88 \pm 0.16$,
which is comparable with the resolution of the temperature bins.
Most DEM codes have difficulty to invert
a sharp high-temperature cutoff, as it was simulated in the models A-C,
while the accuracy increases to $T_w^{fit}/T_w^{sim} = 0.94 \pm 0.12$ 
for model D alone, where the high-temperature cutoff is more gradual.}

\medskip
\item{\underbar{\bf DEM peak emission measure $EM_p$:} The DEM peak
emission measure has been determined with an accuracy of 
$EM_p^{fit}/EM_p^{sim} = 0.64 \pm 0.24$ for the 11 codes and 4 models
(fourth column of Table 3).
We find the most irregular behavior for the Regularized Inversion (RI) 
method, which fails to retrieve the $EM_p$ parameter for models B and C
by far. This may indicate a convergence problem of the RI code.}

\medskip
\item{\underbar{\bf Total DEM emission measure $EM_t$:} The total
(temperature-intergrated) emission measure has been determined with
an accuracy of $EM_t^{fit}/EM_t^{sim} = 0.85 \pm 0.27$ for the 11 codes 
and 4 models (fifth column of Table 3).  
We find the most irregular behavior for the Regularized Inversion (RI) 
method and the Monte Carlo Markov chain (MC) method, which fail to 
retrieve the total emission measure for most cases, even though the fluxes
are fitted extremely well (within $\lapprox 1-2\%$ for the MC method).
The low $\chi^2$-values indicate that the MC method over-fits the data.} 

\medskip
\item{\underbar{\bf Multi-thermal energy $E_{th}$:} The multi-thermal
energy, which is integrated over the entire temperature range of the DEM
(Eq.~7) is matched with an accuracy of $E_{th}^{fit}/E_{th}^{sim} = 
1.2 \pm 0.4$ for the 11 codes and 4 models (sixth column in Table 3).
The MC method yields a poorer accuracy, i.e., $E_{th}^{fit}/E_{th}^{sim} 
= 0.4 \pm 0.1$.}

\medskip
\item{\underbar{\bf Flux ratios $q_F$:} The ratio of the fitted to the
simulated fluxes yields a measure of how well the best-fit model represents
the simulated data. We list the mean flux ratios in Table 3 (eighth column), 
which are averaged from 6 AIA fluxes for most methods, and $\sim 10-20$ 
flux values 
for the methods using EVE data (ER, EG). We see that most of the obtained
ratios have a mean near unity, within a few percent. The only exception
we find is the RI method applied to models B and C, indicating a convergence 
problem. Tendencies of over-fitting are noted for the MCMC and ER
method, based on the untypically small flux errors of $\le 1\%$
(Fig.~9) and large number of degrees of freedom (Table 1).}

\medskip
\item{\underbar{\bf Goodness-of-fit $\chi^2$:} The $\chi^2$ is a 
goodness-of-fit criterion based on the estimated uncertainties, which are
simulated here with 10\% random noise added to a noise-free model in 30
different representations. We list the
means and standard deviations of the $\chi^2$-values in Table 3. It is
instructive to review each code separately in order to judge their overall
adequacy and accuracy. The most accurate codes with a mean value of
$0.5 \lapprox \chi^2 \lapprox 2.0$ are the G1, G2, LS, SS, HX, PS, EG, and ER
codes. There are slight systematic differences, for instance a bi-Gaussian
DEM function (G2) yields in all 4 models a better fit than a single Gaussian
model (G1), which may be a bit fortuitous here, since the simulated DEMs are 
highly asymmetric and thus can naturally be better represented with an
asymmetric DEM function. Mavericks are the MC code ($\chi^2=0.19 \pm 0.09$), 
which appears to overfit the noise, and the RI code ($\chi^2 \approx 150$), 
which appears to have convergence problems. }
\end{enumerate}

\medskip
In summary, the exercise that we conducted here reveals how accurately
we can retrieve physical parameters from AIA, EVE, and RHESSI fluxes, 
and to what degree
DEM inversions are multi-valued or ambiguous, in particular for the 4 
models chosen here. Perhaps the most important physical parameter is the thermal
energy, which we demonstrated can be retrieved within a factor of
$E_{th}^{fit}/E_{th}^{sim} = 1.2 \pm 0.4$ with all 11 codes tested here. 
A large statistical study on thermal energies using AIA data has been 
calculated for $\approx 400$ flare 
events recently (Aschwanden et al.~2015), for which we 
obtain an estimate of the absolute uncertainty here.

For the models tested here, DEM inversions
with AIA data alone yield apparently a comparable accuracy as 
multi-instrument data, such as a combination of EVE with RHESSI or GOES data.
However, we note that all four of our models do not include significant 
high-temperature emission ($\log T \gapprox 7.3$) that would be outside the 
AIA range of sensitivity. For extremely hot flares with DEM peak temperatures of
$\log(T) > 7.3$, DEM modeling should include high-temperature sensitivity
as available from RHESSI and GOES data, as it is done here in conjunction
with EVE data (in the ER and EG method).

The usage of AIA data eliminates uncertainties due 
to cross-calibration, but may underestimate absolute uncertainties of the 
AIA instrument. It is therefore gratifying to see that all three instrument
combinations (EVE+RHESSI, EVE+GOES, AIA) yield equally accurate results 
for the inverted DEM distribution function. Specifically, the
AIA spatial synthesis method, the EVE+GOES method, and the EVE+RHESSI 
method yield the most consistent and accurate results, regardless of
the complex shape of the simulated DEM function.

\acknowledgements 
We dedicate this work to Roger J. Thomas (deceased on May 19, 2015),
with whom several of the authors worked at NASA/GSFC,
an internationally recognized expert in EUV spectrography and GOES 
calibration, which is used in this work here. 
We thank the referee for insightul and constructive comments.
We acknowledge useful discussions with Marc Cheung, Paola Testa, and 
Amy Winebarger. This work is partially supported by NASA under contract 
NNG04EA00C for the SDO/AIA instrument. AC, JMM, and HPW were partially 
supported by NASA grant NNX12AH48G and NASA contracts NAS5-98033 and 
NAS5-02140.

\bibliography{refs}	

\section*{References} 

\def\ref#1{\par\noindent\hangindent1cm {#1}}

\small
\ref{Aschwanden, M.J., Nightingale, R.W., and Boerner, P. 2007, ApJ 656, 577.}
\ref{Aschwanden, M.J. and Tsiklauri, D. 2009, ApJSS 185, 171.}
\ref{Aschwanden, M.J. and Boerner, P., 2011, ApJ 732, 81.}
\ref{Aschwanden, M.J., Boerner, P., Schrijver, C.J., and 
	Mala\-nu\-shen\-ko, A. 2013, Solar Phys. 283, 5.}
\ref{Aschwanden, M.J. and Shimizu, T. 2013, ApJ 776, 132.}
\ref{Aschwanden, M.J., Boerner, P., Ryan, D., Caspi, A., McTiernan, J.M.,
	and Warren, H.P.  2015, ApJ 802, 53 (20pp).}
\ref{Bevington, P.R. and Robinson, D.K.
        1992, {\sl Data reduction and error analysis for the 
	physical sciences}, McGraw Hill: Boston.}
\ref{Boerner, P., Edwards, C., Lemen, J., Rausch, A., Schrijver, C.,
        Shine, R., Shing, L., Stern, R., et al.~2012, Solar Phys. 275, 41.}
\ref{Boerner, P., Testa, P., Warren, H., Weber, M.A., and Schrijver, C.J.
	2014, Solar Phys. 289, 2377.}
\ref{Brosius, J.W., Davila, J.M., Thomas, R.J., and Monsignori-Fossi, B.C.
	1996, ApJS 106, 143.}
\ref{Brown, J.C., Emslie, A.G., and Holman, G.D. et al.~2006, APJ 643, 523.} 
\ref{Caspi, A., Krucker, S., and Lin, R.P. 2014a, ApJ 781, 43.}
\ref{Caspi, A., and Lin, R. P. 2010, ApJL 725, L161}
\ref{Caspi, A., McTiernan, J.M., and Warren, H.P. 2014b, ApJL 788, L31.}
\ref{Cheung, M., Boerner, P., Schrijver, C.M., Testa, P., Chen, F., Peter, H.,
	and Mala\-nu\-shen\-ko, A. 2015, ApJ 807, 143.}
\ref{Craig, I.J.D. and Brown, J.C. 1976, A\&A 49, 239.}
\ref{Garcia, H.A. 1994, Solar Phys. 154, 275.}
\ref{Guennou, C., Auchere, F., Soubrie, E., Bocchialini, K., 
	Parenti, S., and Barbey, N. 2012, ApJSS 203, 26.}
\ref{Golub, L., DeLuca, E., Austin, G., et al. 2007, Sol.Phys. 243, 63.} 
\ref{Feldman, U., Mandelbaum, P., Seely, J.F., Doschek, G.A., and Gursky H.
	1992, ApJSS 81, 387.}
\ref{Inglis, A.R. and Christe, S. 2014, ApJ 789, 116.}
\ref{Hannah, I.G. and Kontar, E.P. 2012, A\&A 539, A146.}
\ref{Hanser, F.A. and Sellers, F.B. 1996, SPIE 2812, 344.}
\ref{Hurford, G.J., Schmahl, E.J., Schwartz, R.A., et al. 2002, Sol.Phys. 210, 61}
\ref{Judge, P.G., Hubeny, V., and Brown, J.C. 1997, ApJ 475, 275.}
\ref{Judge, P.G. 2010, ApJ 708, 1238.}
\ref{Kashyap, V. and Drake, J.J. 1998, ApJ 503, 450.}
\ref{Landi, E. and Klimchuk, J.A. 2010, ApJ 723, 320.}
\ref{Landi, E., Reale, F., and Testa, P. 2012, A\&A 538, A111.}
\ref{Lemen, J.R., Title, A.M., Akin, D.J., Boerner, P.F.,
        Chou, C., Drake, J.F., Duncan, D.W., Edwards, C.G., et al.
        2012, Solar Phys. 275, 17.}
\ref{Lin, R.P., Dennis, B.R., Hurford, G.J., Smith, D.M. et al.
	2002, Solar Phys. 210, 3.}
\ref{Massone, A.M., Emslie, A.G., Kontar, E.P. et al.~2004, ApJ 613, 1233.} 
\ref{Mazzotta, P., Mazzitelli, G., Colafrancesco, S., and Vittorio, N.
	1998, A\&ApS 133, 403.}
\ref{McTiernan, J.M. 2009, ApJ 697, 94}
\ref{Monsignori-Fossi, B.C. and Landini, M. 1992, Mem. Soc. Astron. Ital.
	63, 767.}
\ref{O'Dwyer, B., DelZanna, G., Mason, H.E., Weber, M.A., Tripathi, D.
 	2010, A\&A 521, A21.}
\ref{Parenti, S., Bromage, B.J.I., Poletto, G., Noci, G., Raymond, J.C>,
	and Bromage, G.E. 2000, A\&A 363, 800.}
\ref{Piana, M., Massone, A.M., Kontar, E.P. et al.~2003, ApJ 595, L127.}
\ref{Pesnell, W.D., Thompson, B.J., and Chamberlin P.C. 2012,
	Solar Phys. 275, 3.}
\ref{Phillips, K.J.H. 2004, ApJ 605, 921}
\ref{Phillips, K.J.H., Chifor, C., and Dennis,B. 2006,
 	ApJ 647, 1480.}
\ref{Phillips, K.J.H. 2008, 
 	Astron.\& Astroph. 490, 828.}
\ref{Prato, M., Piana, M., Brown, J.C. et al.~2006, Solar Phys. 237, 61.}
\ref{Press, W.H., Flannery, B.P., Teukolsky, S.A., and Vetterling, W.T.
 	1986, {\sl Numerical recipes, The Art of Scientific Computing},
	Cambridge University Press: Cambridge.} 
\ref{Rosner, R., Tucker, W.H., and Vaiana, G.S. 1978, ApJ 220, 643.}
\ref{Ryan, D.F., Milligan, R.O., Gallagher, P.T., Dennis, B.R., Tolbert, A.K.,
	Schwartz, R.A., and Young, C.A. 2012, ApJSS 202, 11.}
\ref{Ryan, D.F., O'Flannagain, A.M., Aschwanden, M.J., and Gallagher, P.T.
	2014, Solar Physics, online-first, DOI 10.1007/s11207-014-0492-z.}
\ref{Scullion, E., Rouppe van der Voort, L., Wedemeyer, S., and Antolin, P.
	2014, ApJ 797, 36.}
\ref{Smith, D.M., Lin, R.P., Turin, P., et al. 2002, Sol.Phys. 210, 33}
\ref{Thomas, R.J., Crannell, C.J., and Starr, R. 1985, Solar Phys. 95, 323.}
\ref{Teriaca, L., Warren, H.P., and Curdt, W. 2012, ApJL 754, L40.}
\ref{Testa, P., De Pontieu, B., Martinez-Sykora J., Hansteen, V., 
	and Carlsson, M. 2012a, ApJ 758, 54.}
\ref{Testa, P., Drake, J.J., and Landi, E. 2012b, ApJ 745, 111.}
\ref{Warren, H.P. 2005 ApJSS 157, 147.}
\ref{Warren, H.P., Mariska, J.T., and Doschek, G.A. 2013, ApJ 770, 116.}
\ref{Warren, H.P. 2014, ApJ 786, L2.}
\ref{Weber, M.A., DeLuca, E.E,, Golub, B., and Sette, A.L. 2004,
	in {\sl Multi-Wavelength Investigations of Solar Activity},
	(eds. A.V. Stepanov, E.E. Benevolenskaya,, and A.G. Kosovichev),
	IAU Symp. 223, 321.}
\ref{White, R.J., Crannell, C.J., and Starr, R. 2005, Solar Phys. 227, 231.}
\ref{Winebarger, A.R., Warren, H.P., Schmelz, J.T., and Cirtain, J.
	2012, ApJL 746, L17.}
\ref{Woods, T.N., Eparvier, F.G., Hock, R., et al.~2012, Solar Phys. 275, 115.}

\end{article}
\end{document}